%% file: SMP-17-014_temp.tex
\pdfoutput=1

\documentclass[11pt,twoside,a4paper,cmspaper,final,collab]{cms-tdr}
\def\svnVersion{493142P}\def\cmsCernNoTag{CERN-EP-2018-282}\def\cmsCernDate{\today}\def\cmsMessage{Published in the European Physical Journal C as \href{http://dx.doi.org/10.1140/epjc/s10052-019-6752-1}{\doi{10.1140/epjc/s10052-019-6752-1.}}}

\begin{document}\cmsNoteHeader{SMP-17-014}

\hyphenation{had-ron-i-za-tion}
\hyphenation{cal-or-i-me-ter}
\hyphenation{de-vices}
\RCS$HeadURL: svn+ssh://svn.cern.ch/reps/tdr2/papers/SMP-17-014/trunk/SMP-17-014.tex $
\RCS$Id: SMP-17-014.tex 489421 2019-02-19 18:28:16Z alverson $

\newlength\cmsFigWidth
\ifthenelse{\boolean{cms@external}}{\setlength\cmsFigWidth{0.85\columnwidth}}{\setlength\cmsFigWidth{0.4\textwidth}}
\ifthenelse{\boolean{cms@external}}{\providecommand{\cmsLeft}{upper\xspace}}{\providecommand{\cmsLeft}{left\xspace}}
\ifthenelse{\boolean{cms@external}}{\providecommand{\cmsRight}{lower\xspace}}{\providecommand{\cmsRight}{right\xspace}}

\newcommand{\asmz}{\ensuremath{\alpS(m_\PZ)}\xspace}
\newcommand{\sqrts}{\ensuremath{\sqrt{s}}}
\newcommand{\intLumi}{\ensuremath{\Lumi_{\text{int}}}\xspace}
\newcommand{\Nsel}{\ensuremath{N_{\text{sel}}}\xspace}

\newcommand{\wc}{\ensuremath{\PW{+}\cPqc}\xspace}
\newcommand{\sigmawc}{\ensuremath{\sigma(\wc)}\xspace}
\newcommand{\dsigmawc}{\ensuremath{\rd\sigmawc}\xspace}
\newcommand{\wpluscbar}{\ensuremath{\PWp{+}\cPaqc}\xspace}
\newcommand{\swpluscbar}{\ensuremath{\sigma(\wpluscbar)}\xspace}
\newcommand{\dswpluscbar}{\ensuremath{\rd\swpluscbar}\xspace}
\newcommand{\wminusc}{\ensuremath{\PWm{+}\cPqc}\xspace}
\newcommand{\swminusc}{\ensuremath{\sigma(\wminusc)}\xspace}
\newcommand{\dswminusc}{\ensuremath{\rd\swminusc}\xspace}

\newcommand{\Dstar}{\ensuremath{\PDstpm}\xspace}
\newcommand{\deltam}{\ensuremath{\Delta m(\PD^*, \PDz)}\xspace}
\newcommand{\wDstar}{\ensuremath{\PW{+}\PDstpm}\xspace}
\newcommand{\swDstar}{\ensuremath{\sigma(\PW{+}\PDstpm)}\xspace}
\newcommand{\wpDstar}{\ensuremath{\PWp{+}\PD^*(2010)^-}\xspace}
\newcommand{\swpDstar}{\ensuremath{\sigma(\wpDstar)}\xspace}
\newcommand{\wmDstar}{\ensuremath{\PWm{+}\PD^*(2010)^+}\xspace}
\newcommand{\swmDstar}{\ensuremath{\sigma(\wmDstar)}\xspace}

\newcommand{\ptCharm}{\ensuremath{\pt^\cPqc}\xspace}
\newcommand{\ptMuon}{\ensuremath{\pt^{\PGm}}\xspace}
\newcommand{\etaMuon}{\ensuremath{\abs{\eta^{\PGm}}}\xspace}
\newcommand{\detaMuon}{\ensuremath{\rd\etaMuon}\xspace}
\newcommand{\etaMuonMinMax}{\ensuremath{[\abs{\eta^{\Pgm}_{\text{min}}}, \abs{\eta^{\Pgm}_{\text{max}}}] }\xspace}
\newcommand{\OSminusSS}{\ensuremath{\text{OS-SS}}\xspace}

\newcommand{\mWsquared}{\ensuremath{m^2_{\PW}}\xspace}

\newcommand{\ptDstar}{\ensuremath{\pt^{\PD^*}}\xspace}
\newcommand{\etaDstar}{\ensuremath{\abs{\eta^{\PD^*}}}\xspace}
\newcommand{\PiSlowpm}{\ensuremath{\Pgp_{\text{slow}}^{\pm}}\xspace}
\newcommand{\Kaonmp}{\ensuremath{\PK^{\mp}}\xspace}
\newcommand{\ptTrack}{\ensuremath{\pt^{\text{track}}}\xspace}

\newcommand{\renormScale}{\ensuremath{\mu_\mathrm{r}}\xspace}
\newcommand{\factScale}{\ensuremath{\mu_\mathrm{f}}\xspace}
\newcommand{\chiSquare}{\ensuremath{\chi^2}\xspace}

\newcommand{\Wjets}{\PW{+}jets\xspace}

\providecommand{\cmsTable}[1]{\resizebox{\textwidth}{!}{#1}}
\newlength\cmsTabSkip\setlength{\cmsTabSkip}{1ex}

\cmsNoteHeader{SMP-17-014}
\title{Measurement of associated production of a $\PW$ boson and a charm quark in proton-proton collisions at $\sqrt{s} = 13\TeV$}
\titlerunning{Associated production of a \PW{} boson and a charm quark}

\date{\today}

\abstract{
Measurements are presented of associated production of a $\PW$ boson and a charm quark ($\PW+\cPqc$) in proton-proton collisions at a center-of-mass energy of 13\TeV.
The data correspond to an integrated luminosity of 35.7\fbinv collected by the CMS experiment at the CERN LHC.
The $\PW$ bosons are identified by their decay into a muon and a neutrino.
The charm quarks are tagged via the full reconstruction of $\PDstpm$ mesons that decay via $\PDstpm \to \PDz + \Pgppm \to \PK^{\mp} +  \Pgppm + \Pgppm$.
A cross section is measured in the fiducial region defined by the muon transverse momentum $\pt^{\PGm} > 26\GeV$, muon pseudorapidity $\abs{\eta^{\PGm}} < 2.4$, and charm quark transverse momentum $\pt^{\cPqc} > 5\GeV$.
The inclusive cross section for this kinematic range is $\sigma(\PW+\cPqc)=1026\pm 31\stat\substack{+76\\-72}\syst\unit{pb}$.
The cross section is also measured differentially as a function of the pseudorapidity of the muon from the $\PW$ boson decay.
These measurements are compared with theoretical predictions and are used to probe the strange quark content of the proton.
}

\hypersetup{
pdfauthor={CMS Collaboration},%
pdftitle={W plus Charm quark production at 13 TeV},%
pdfsubject={CMS},
pdfkeywords={CMS, standard model physics, proton structure}}

\maketitle

\section{Introduction}
Precise knowledge of the structure of the proton, expressed in terms of parton distribution functions (PDFs), is important for interpreting results obtained in proton-proton (\Pp\Pp) collisions at the CERN LHC.
The PDFs are determined by comparing theoretical predictions obtained at a particular order in perturbative quantum chromodynamics (pQCD) to experimental measurements.
The precision of the PDFs, which affects the accuracy of the theoretical predictions for cross sections at the LHC, is determined by the uncertainties of the experimental measurements used, and  by the limitations of the available theoretical calculations.
The flavor composition of the light quark sea in the proton and, in particular, the understanding of the strange quark distribution is important for the measurement of the $\PW$ boson mass at the LHC~\cite{Aaboud:2017svj}.
Therefore, it is of great interest to determine the strange quark distribution with improved precision.

Before the start of LHC data taking, information on the strange quark content of the nucleon was obtained primarily from charm production in (anti)neutrino-iron deep inelastic scattering (DIS) by the NuTeV~\cite{Goncharov:2001qe}, CCFR~\cite{Bazarko:1994tt}, and NOMAD~\cite{Samoylov:2013xoa} experiments.
In addition, a direct measurement of inclusive charm production in nuclear emulsions was performed by the CHORUS experiment~\cite{KayisTopaksu:2011mx}.
At the LHC, the production of $\PW$ or $\PZ$ bosons, inclusive or associated with charm quarks, provides an important input for tests of the earlier determinations of the strange quark distribution.
The measurements of inclusive $\PW$ or $\PZ$ boson production at the LHC, which are indirectly sensitive to the strange quark distribution, were used in a QCD analysis by the ATLAS experiment, and an enhancement of the strange quark distribution with respect to other measurements was observed~\cite{Aad:2012sb}.

\begin{figure}[h]
	\centering
	\includegraphics[width=0.3\textwidth]{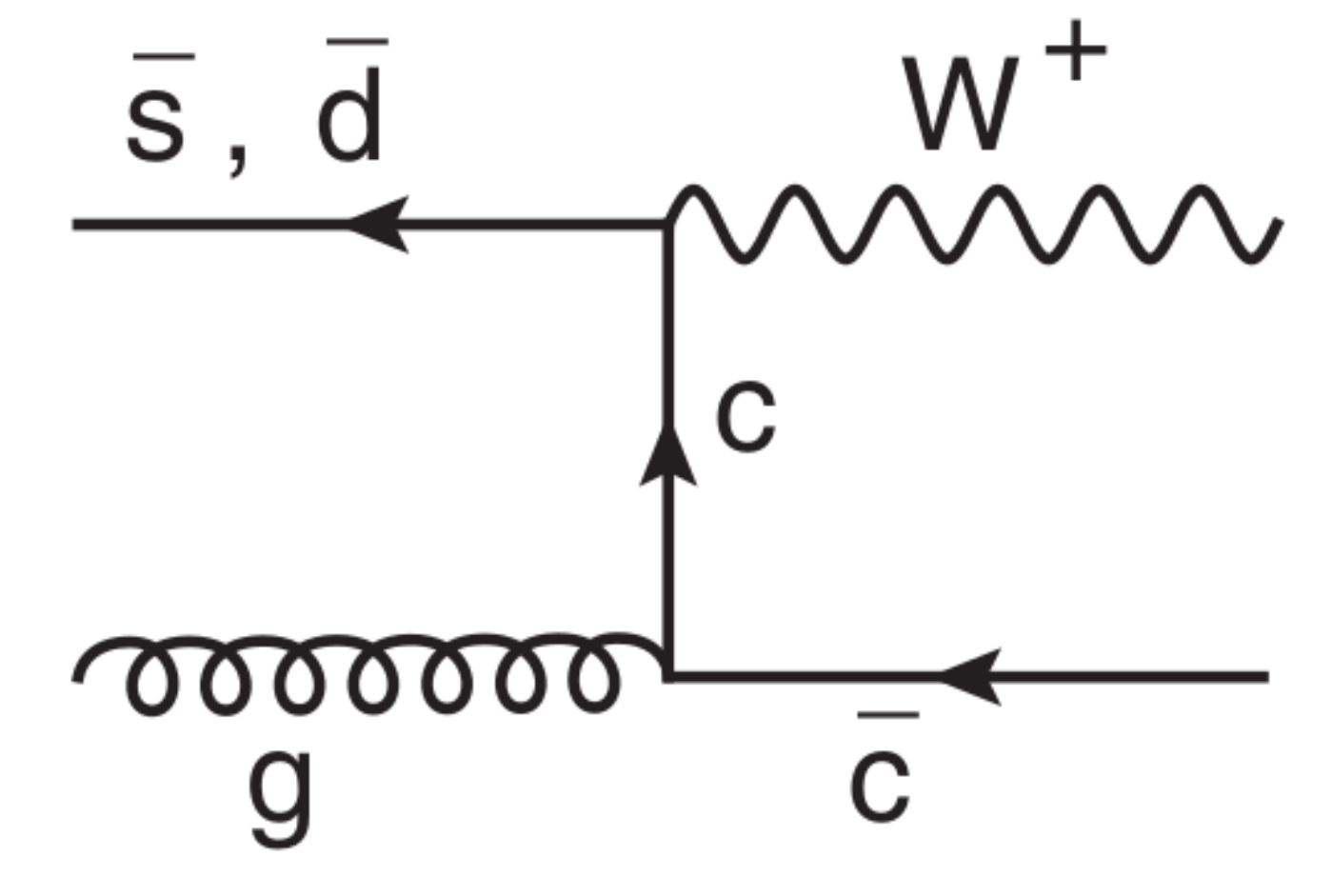}
	\includegraphics[width=0.3\textwidth]{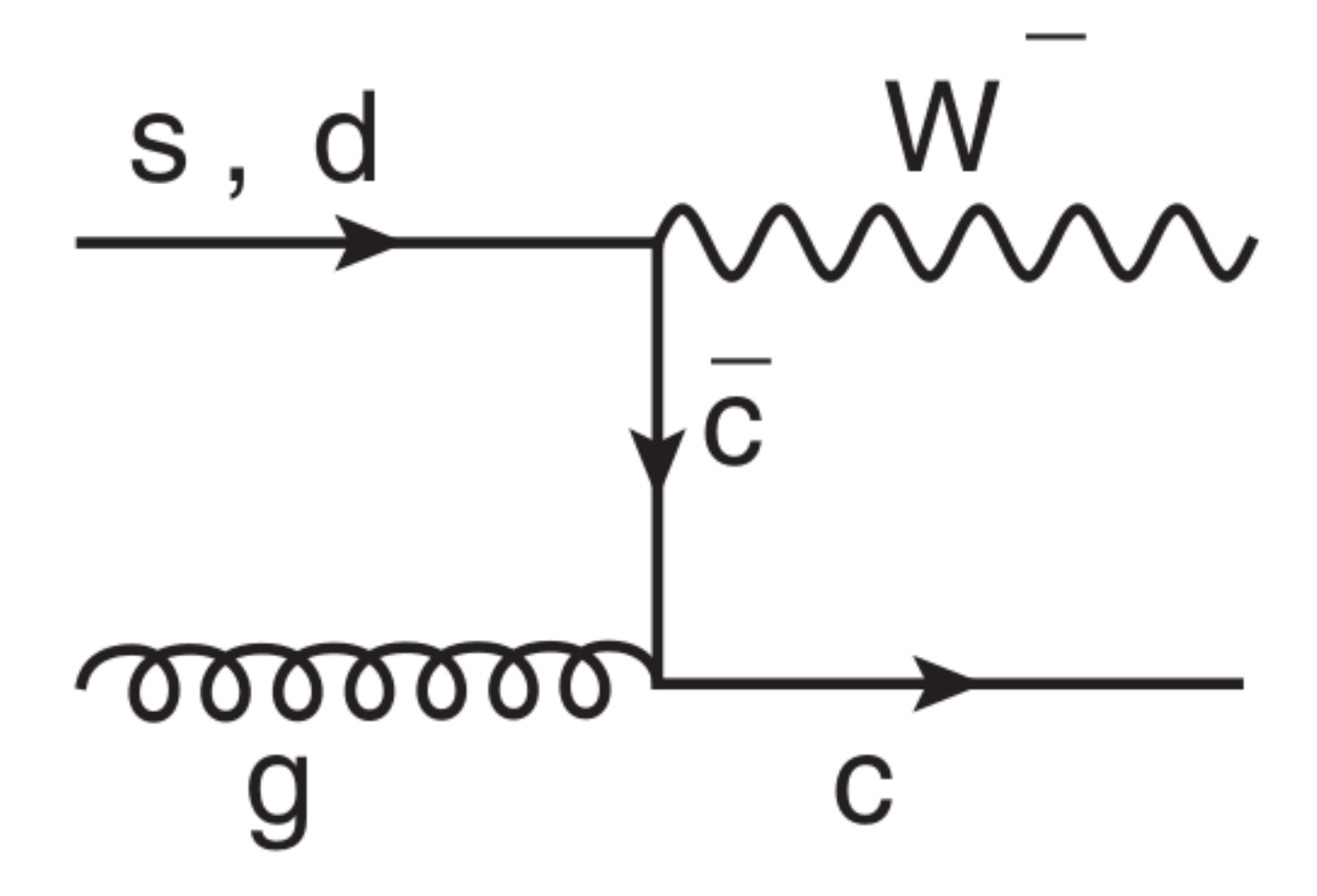}
	\caption{Dominant contributions to \wc production at the LHC at leading order in pQCD.}
	\label{Wc_Feynman}
\end{figure}

The associated production of $\PW$ bosons and charm quarks in pp collisions at the LHC probes the strange quark content of the proton directly through the leading order (LO) processes $\cPg + \cPaqs \to \wpluscbar$ and $\cPg + \cPqs \to \wminusc$, as shown in Fig.~\ref{Wc_Feynman}.
The contribution of the Cabibbo-suppressed processes $\cPg+ \cPaqd\to \wpluscbar$ and $\cPg+\cPqd \to \wminusc$ amounts to only a few percent of the total cross section.
Therefore, measurements of associated \wc production in pp collisions provide valuable insights into the strange quark distribution of the proton.
Furthermore, these measurements allow important cross-checks of the results obtained in the global PDF fits using the DIS data and measurements of inclusive $\PW$ and $\PZ$ boson production at the LHC.

Production of \wc in hadron collisions was first investigated at the Tevatron~\cite{Aaltonen:2007dm,Aaltonen:2012wn,Abazov:2008qz}.
The first measurement of the cross section of \wc production in $\Pp\Pp$ collisions at the LHC was performed by the CMS Collaboration at a center-of-mass energy of $\sqrts = 7\TeV$ with an integrated luminosity of 5\fbinv~\cite{Chatrchyan:2013uja}.
This measurement was used for the first direct determination of the strange quark distribution in the proton at a hadron collider~\cite{Chatrchyan:2013mza}.
The extracted strangeness suppression with respect to $\cPaqu$ and $\cPaqd$ quark densities was found to be in agreement with measurements in neutrino scattering experiments.
The cross section for \wc production was also measured by the ATLAS experiment at $\sqrts = 7\TeV$~\cite{Aad:2014xca} and used in a QCD analysis, which supported the enhanced strange quark content in the proton suggested by the earlier ATLAS analysis~\cite{Aad:2012sb}.
A subsequent joint QCD analysis~\cite{Alekhin:2014sya} of all available data that were sensitive to the strange quark distribution demonstrated consistency between the \wc measurements by the ATLAS and CMS Collaborations.
In Ref.~\cite{Alekhin:2014sya}, possible reasons for the observed strangeness enhancement were discussed.
Recent results of an ATLAS QCD analysis~\cite{Aaboud:2016btc}, including measurements of inclusive $\PW$ and $\PZ$ boson production at $\sqrts = 7\TeV$, indicated an even stronger strangeness enhancement in disagreement with all global PDFs.
In Ref.~\cite{Alekhin:2017olj}, possible reasons for this observation were attributed to the limitations of the parameterization used in this ATLAS analysis~\cite{Aaboud:2016btc}.
The associated production of a $\PW$ boson with a jet originating from a charm quark is also studied in the forward region by the LHCb experiment~\cite{Aaij:2015cha}.

In this paper, the cross section for \wc production is measured in pp collisions at the LHC at $\sqrts = 13\TeV$ using data collected by the CMS experiment in 2016 corresponding to an integrated luminosity of 35.7\fbinv.
The $\PW$ bosons are selected via their decay into a muon and a neutrino.
The charm quarks are tagged by a full reconstruction of the charmed hadrons in the process $\cPqc \to \PDstpm \to \PDz + \PiSlowpm \to \Kaonmp + \Pgppm + \PiSlowpm$, which has a clear experimental signature.
The pion originating from the $\PDstpm$ decay receives very little energy because of the small mass difference between $\PDstpm$ and $\PDz$(1865) and is therefore denoted a ``slow" pion $\PiSlowpm$.
Cross sections for \wDstar production are measured within a selected fiducial phase space.
The \wc cross sections are compared with theoretical predictions at next-to-leading order (NLO) QCD, which are obtained with {\MCFM 6.8}~\cite{Campbell:1999ah,Campbell:2010ff,Campbell:2012uf}, and are used to extract the strange quark content of the proton.

This paper is organized as follows.
The CMS detector is briefly described in Section~\ref{Detector}.
The data and the simulated samples are described in Section~\ref{Samples}.
The event selection is presented in Section~\ref{EventSelection}.
The measurement of the cross sections and the evaluation of systematic uncertainties are discussed in Section~\ref{CrossSection}.
The details of the QCD analysis are described in Section~\ref{sec:QCD}.
Section~\ref{Summary} summarizes the results.

\section{The CMS detector}
\label{Detector}
The central feature of the CMS apparatus is a superconducting solenoid of 6\unit{m} internal diameter, providing a magnetic field of 3.8\unit{T}.
Within the solenoid volume are a silicon pixel and strip tracker, a lead tungstate crystal electromagnetic calorimeter (ECAL), and a brass and scintillator hadron calorimeter (HCAL), each composed of a barrel and two endcap sections.
Forward calorimeters extend the pseudorapidity coverage provided by the barrel and endcap detectors.
Muons are detected in gas-ionization chambers embedded in the steel flux-return yoke outside the solenoid.

The silicon tracker measures charged particles within the pseudorapidity range $\abs{\eta} < 2.5$. It consists of 1440 silicon pixel and 15\,148 silicon strip detector modules.
For nonisolated particles of $1 < \pt < 10\GeV$ and $\abs{\eta} < 1.4$, the track resolutions are typically 1.5\% in \pt and 25--90 (45--150)\mum in the transverse (longitudinal) impact parameter \cite{Chatrchyan:2014fea}.
The reconstructed vertex with the largest value of summed physics-object $\pt^2$ is taken to be the primary $\Pp\Pp$ interaction vertex.
The physics objects are the jets, clustered using the jet finding algorithm~\cite{Cacciari:2008gp,Cacciari:2011ma} with the tracks assigned to the vertex as inputs, and the associated missing transverse momentum, taken as the negative vector sum of the \pt of those jets.
Muons are measured in the pseudorapidity range $\abs{\eta} < 2.4$, with detection planes made using three technologies: drift tubes, cathode strip chambers, and resistive-plate chambers.
The single muon trigger efficiency exceeds 90\% over the full $\eta$ range, and the efficiency to reconstruct and identify muons is greater than 96\%.
Matching muons to tracks measured in the silicon tracker results in a relative transverse momentum resolution, for muons with \pt up to 100\GeV, of 1\% in the barrel and 3\% in the endcaps.
The \pt resolution in the barrel is better than 7\% for muons with \pt up to 1\TeV~\cite{Sirunyan:2018fpa}.
A more detailed description of the CMS detector, together with a definition of the coordinate system used and the relevant kinematic variables, can be found in Ref.~\cite{Chatrchyan:2008zzk}.

\section{Data and Monte Carlo samples and signal definition}
\label{Samples}

Candidate events for the muon decay channel of the $\PW$ boson are selected by a muon trigger~\cite{Khachatryan:2016bia} that requires a reconstructed muon with $\ptMuon > 24\GeV$.
The presence of a high-\pt neutrino is implied by the missing transverse momentum, $\ptvecmiss$, which is defined as the negative vector sum of the transverse momenta of the reconstructed particles.

Muon candidates and $\ptvecmiss$ are reconstructed using the particle-flow (PF) algorithm~\cite{Sirunyan:2017ulk}, which reconstructs and identifies each individual particle with an optimized combination of information from the various elements of the CMS detector.
The energy of photons is obtained directly from the ECAL measurement.
The energy of electrons is determined from a combination of the electron momentum at the primary interaction vertex determined by the tracking detector, the energy of the corresponding ECAL cluster, and the energy sum of all bremsstrahlung photons spatially compatible with originating from the electron track.
The muon momentum is obtained from the track curvature in both the tracker and the muon system, and identified by hits in multiple stations of the flux-return yoke.
The energy of charged hadrons is determined from a combination of their momentum measured in the tracker and the matching ECAL and HCAL energy deposits, corrected for both zero-suppression effects and the response function of the calorimeters to hadronic showers.
Finally, the energy of neutral hadrons is obtained from the corresponding corrected ECAL and HCAL energy.

{\tolerance=800
The $\PDstpm$ meson candidates are reconstructed from tracks formed by combining the measurements in the silicon pixel and strip detectors through the CMS combinatorial track finder~\cite{Chatrchyan:2014fea}.  \par}

The signal and background processes are simulated using Monte Carlo (MC) generators to estimate the acceptance and efficiency of the CMS detector.
The corresponding MC events are passed through a detailed \GEANTfour~\cite{Agostinelli:2002hh} simulation of the CMS detector and reconstructed using the same software as the real data.
The presence of multiple pp interactions in the same or adjacent bunch crossing (pileup) is incorporated by simulating additional interactions (both in-time and out-of-time with respect to the hard interaction) with a vertex multiplicity that matches the distribution observed in data.
The simulated samples are normalized to the integrated luminosity of the data using the generated cross sections.
To simulate the signal, inclusive \Wjets events are generated with \MGvATNLO~(v2.2.2)~\cite{Alwall:2014hca} using the NLO matrix elements, interfaced with {\PYTHIA}8~(8.2.12)~\cite{Sjostrand:2007gs} for parton showering and hadronization.
A matching scale of 10\GeV is chosen, and the \textsc{FxFx} technique~\cite{Frederix:2012ps} is applied for matching and merging.
The factorization and renormalization scales, $\renormScale^2$ and $\factScale^2$, are set to $\renormScale^2=\factScale^2 = m^2_{\PW} + p^2_{\mathrm{T},\PW}$.
The proton structure is described by the NNPDF3.0nlo~\cite{Ball:2014uwa} PDF set.
To enrich the sample with simulated \wc events, an event filter that requires at least one muon with $\ptMuon > 20\GeV$ and $\etaMuon < 2.4$, as well as at least one $\PDstpm$ meson, is applied at the generator level.

{\tolerance=800
Several background contributions are considered, which are described in the following.
An inclusive \Wjets event sample is generated using the same settings as the signal events, but without the event filter, to simulate background contributions from \PW~events that do not contain \Dstar mesons.
Events originating from Drell--Yan (DY) with associated jets are simulated with \MGvATNLO~(v2.2.2) with $\renormScale^2$ and $\factScale^2$ set to $m^2_{\PZ} + p^2_{\mathrm{T},\PZ}$.
Events originating from top quark-antiquark pair ($\ttbar$) production are simulated using \POWHEG(v2.0)~\cite{Campbell:2014kua}, whereas single top quark events are simulated using \POWHEG(v2.0)~\cite{Frederix:2012dh,Alioli:2009je} or \POWHEG(v1.0)~\cite{Re:2010bp}, depending on the production channel.
Inclusive production of $\PW \PW$, $\PW \PZ$, and $\PZ \PZ$ bosons and contributions from the inclusive QCD events are generated using {\PYTHIA}8.
The CUETP8M1~\cite{Khachatryan:2015pea} underlying event tune is used in {\PYTHIA}8 for all, except for the $\ttbar$ sample, where the \\ CUETP8M2T4~\cite{CMS-PAS-TOP-16-021} tune is applied. \par}

The dominant background originates from processes like $\cPqu + \cPaqd \to \PWp + \cPg^* \to \PWp + \ccbar$ or $\cPqd + \cPaqu \to \PWm + \cPg^* \to \PWm + \ccbar$, with $\cPqc$ quarks produced in gluon splitting.
In the \wc signal events the charges of the $\PW$ boson and the charm quark have opposite signs.
In gluon splitting, an additional $\cPqc$ quark is produced with the same charge as the $\PW$ boson.
At the generator level, an event is considered as a \wc event if it contains at least one charm quark in the final state.
In the case of an odd number of $\cPqc$ quarks, the $\cPqc$ quark with the highest $\pt$ and a charge opposite to that of the $\PW$ boson is considered as originating from a \wc process, whereas the other $\cPqc$ quarks in the event are labeled as originating from gluon splitting.
In the case of an even number of $\cPqc$ quarks, all are considered to come from gluon splitting.
Events containing both $\cPqc$ and $\cPqb$ quarks are considered to be \wc~events, since $\cPqc$ quarks are of higher priority in this analysis, regardless of their momentum or production mechanism.
Events containing no $\cPqc$ quark and at least one $\cPqb$ quark are classified as $\PW + \cPqb$.
Otherwise, an event is assigned to the $\PW+\cPqu\cPqd\cPqs\Pg$ category.

The contribution from gluon splitting can be significantly reduced using data.
Events with the same charge sign for both the $\PW$ boson and charm quark, which correlates to the charge sign of the \Dstar meson, are background, which is due to gluon splitting.
Since the gluon splitting background for opposite charge pairs is identical, it can be removed by subtracting the same-sign distribution from the signal.
The measurement is performed in the central kinematic range and is not sensitive to the contributions of processes $\cPqc + \cPg \to \PW + \cPqs$ with a spectator charm quark.

For validation and tuning of MC event generators using a Rivet plugin~\cite{Buckley:2010ar}, the $\wDstar$ measurement is performed.
This requires a particle-level definition without constraints on the origin of \Dstar mesons.
Therefore, any contributions from $\PB$ meson decays and other hadrons, though only a few pb, are included as signal for this part of the measurement.

\section{Event selection}
\label{EventSelection}
The associated production of $\PW$ bosons and charm quarks is investigated using events, where $\PW \to \Pgm + \Pagngm$ and the $\cPqc$ quarks hadronize into a $\PDstpm$ meson.
The reconstruction of the muons from the $\PW$ boson decays and of the $\PDstpm$ candidates is described in detail in the following.

\subsection{Selection of \texorpdfstring{\PW}{W} boson candidates}
Events containing a $\PW$ boson decay are identified by the presence of a high-\pt isolated muon and \ptvecmiss.
The muon candidates are reconstructed by combining the tracking information from the muon system and from the inner tracking system~\cite{Sirunyan:2018fpa}, using the CMS particle-flow algorithm.
Muon candidates are required to have $\ptMuon > 26\GeV$, $\etaMuon < 2.4$, and must fulfill the CMS ``tight identification" criteria~\cite{Sirunyan:2018fpa}.
To suppress contamination from muons contained in jets, an isolation requirement is imposed:
\begin{linenomath}
\ifthenelse{\boolean{cms@external}}
{
\begin{multline*}
\frac{1}{\ptMuon} \Biggl[ \sum^{\mathrm{CH}} \pt + \max \Bigl( 0., \sum^{\mathrm{NH}} \pt + \sum^{\mathrm{EM}} \pt - 0.5 \sum^{\mathrm{PU}} \pt  \Bigr) \Biggl] \\
 			\leq 0.15,
\end{multline*}
}
{
\begin{equation}
 \frac{1}{\ptMuon} \left[ \sum^{\mathrm{CH}} \pt + \max \left( 0., \sum^{\mathrm{NH}} \pt + \sum^{\mathrm{EM}} \pt - 0.5 \sum^{\mathrm{PU}} \pt  \right) \right]  \leq 0.15,
\end{equation}
}
\label{MuonIsolation}
\end{linenomath}
where the $\pt$ sum of PF candidates for charged hadrons (CH), neutral hadrons (NH), photons (EM) and charged particles from pileup (PU) inside a cone of radius $\Delta R \leq 0.4$ is used, and the factor 0.5 corresponds to the typical ratio of neutral to charged particles, as measured in jet production~\cite{Sirunyan:2017ulk}.

{\tolerance=800
Events in which more than one muon candidate fulfills all the above criteria are rejected in order to suppress background from DY processes.
Corrections are applied to the simulated samples to adjust the trigger, isolation, identification, and tracking efficiencies to the observed data.
These correction factors are determined through dedicated tag-and-probe studies.  \par}

The presence of a neutrino in an event is assured by imposing a requirement on the transverse mass, which is defined as the combination of $\ptMuon$ and $\ptvecmiss$:
\begin{equation}
	\mT \equiv \sqrt{{2 \, \ptMuon \, \ptvecmiss \, (1 - \cos(\phi_{\Pgm} - \phi_{\ptvecmiss}))}}.
	\label{TransverseMass}
\end{equation}
In this analysis, $\mT > 50\GeV$ is required, which results in  a significant reduction of background.

\subsection{Selection of \texorpdfstring{\PDstpm}{D*(2010)}  candidates}

{\tolerance=800
The \Dstar mesons are identified by their decays $\PDstpm \to \PDz + \PiSlowpm \to \Kaonmp + \Pgppm + \PiSlowpm$ using the reconstructed tracks of the decay products.
The branching fraction for this channel is $2.66 \pm 0.03\%$~\cite{PDG2018}. \par}

{\tolerance=1000
The $\PDz$ candidates are constructed by combining two oppositely charged tracks with transverse momenta $\ptTrack > 1\GeV$, assuming the $\Kaonmp$ and $\Pgppm$ masses.
The $\PDz$ candidates are further combined with a track of opposite charge to the kaon candidate, assuming the $\Pgp^{\pm}$ mass, following the well-established procedure of Ref.~\cite{Nussinov:1975ay,Feldman:1977ir}.
The invariant mass of the $\Kaonmp\Pgppm$ combination is required to be $\abs{m(\Kaonmp\Pgppm) - m(\PDz)} < 35\MeV$, where $m(\PDz) = 1864.8 \pm 0.1\MeV$~\cite{PDG2018}.
The candidate $\Kaonmp$ and $\Pgppm$ tracks must originate at a fitted secondary vertex~\cite{Fruhwirth:1027031} that is displaced by not more than 0.1\unit{cm} in both the $xy$-plane and $z$-coordinate from the third track, which is presumed to be the $\PiSlowpm$ candidate.
The latter is required to have $\ptTrack > 0.35\GeV$ and to be in a cone of $\Delta R \leq 0.15$ around the direction of the $\PDz$ candidate momentum.
The combinatorial background is reduced by requiring the \Dstar transverse momentum $\ptDstar > 5\GeV$ and by applying an isolation criterion $\ptDstar / \sum \pt > 0.2$.
Here $\sum \pt$ is the sum of transverse momenta of tracks in a cone of $\Delta R \leq$ 0.4 around the direction of the \Dstar momentum.
The contribution of \Dstar mesons produced in pileup events is suppressed by rejecting candidates with a $\text{$z$-distance} > 0.2\unit{cm}$ between the muon and the $\PiSlowpm$.
After applying all selection criteria, the contribution of $\PDz$ decays other than $\Kaonmp\Pgppm$ is negligible compared to the uncertainties.
\par }

The \Dstar meson candidates are identified using the mass difference method~\cite{Feldman:1977ir} via a peak in the \deltam distribution.
Wrong-charge combinations with $\PKpm \Pgppm$ pairs in the accepted $\PDz$ mass range mimic the background originating from light-flavor hadrons.
By subtracting the wrong-charge combinations, the combinatorial background in the \deltam distribution is mostly removed.
The presence of nonresonant charm production in the right-charge $\Kaonmp \Pgppm \Pgppm$ combinations introduces a small normalization difference of $\deltam$ distributions for right- and wrong-charge combinations, which is corrected utilizing fits to the ratio of both distributions.

\subsection{Selection of \wc candidates}

An event is selected as a \wc signal if it contains a $\PW$ boson and a $\Dstar$ candidate fulfilling all selection criteria.
The candidate events are split into two categories: with $\PW^\pm +\Dstar$  combinations falling into the same sign (SS) category, and $\PW^\mp + \Dstar$  combinations falling into the opposite sign (OS) category.
The signal events consist of only OS combinations, whereas the $\PW+ \ccbar$ and $\PW+ \bbbar$ background processes produce the same number of OS and SS candidates.
Therefore, subtracting the SS events from the OS events removes the background contributions from gluon splitting.
The contributions from other background sources, such as $ \cPqt \cPaqt$ and single top quark production, are negligible.

\begin{figure}[th]
\centering
	\includegraphics[width=0.48\textwidth]{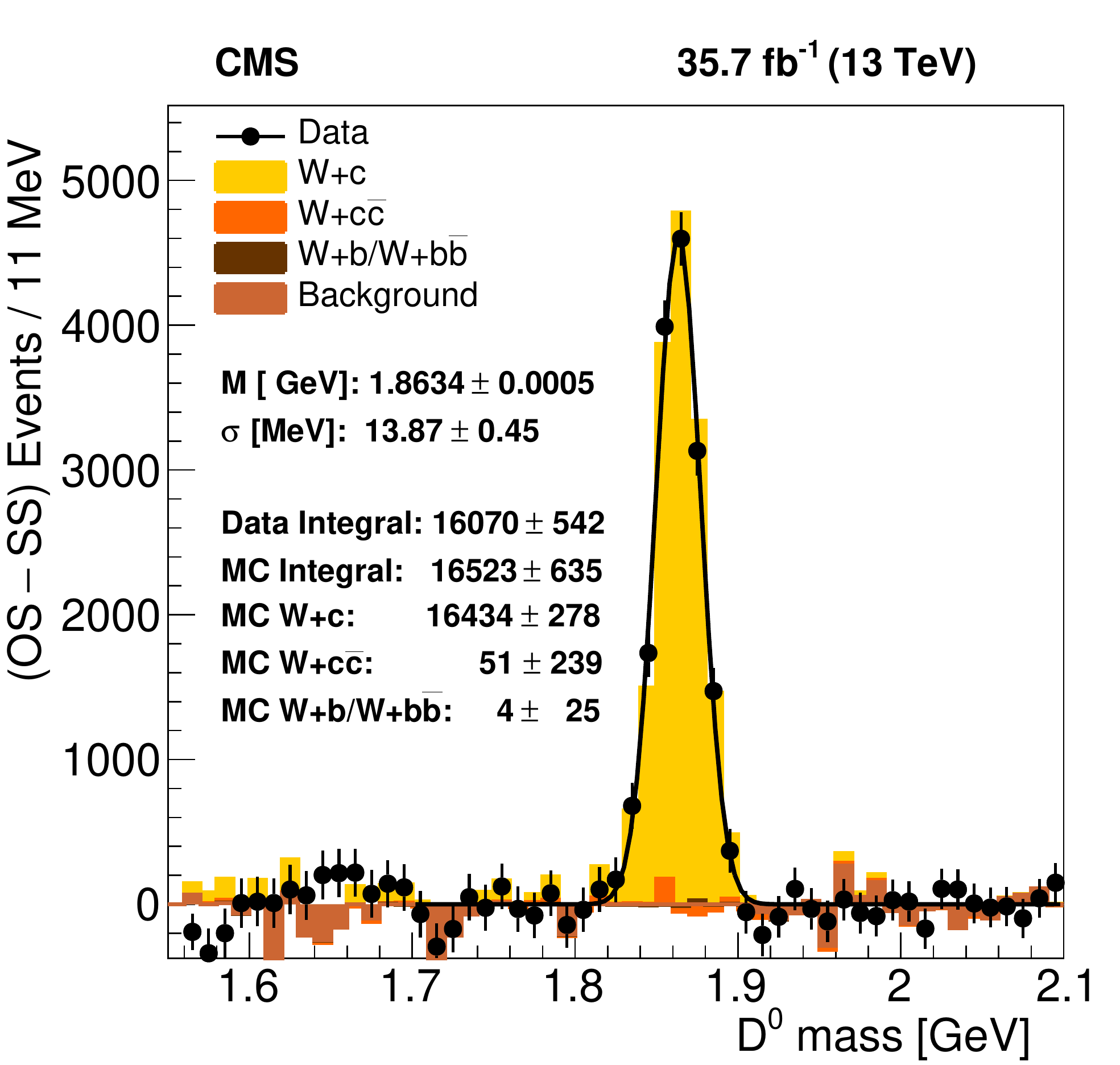}
	\includegraphics[width=0.48\textwidth]{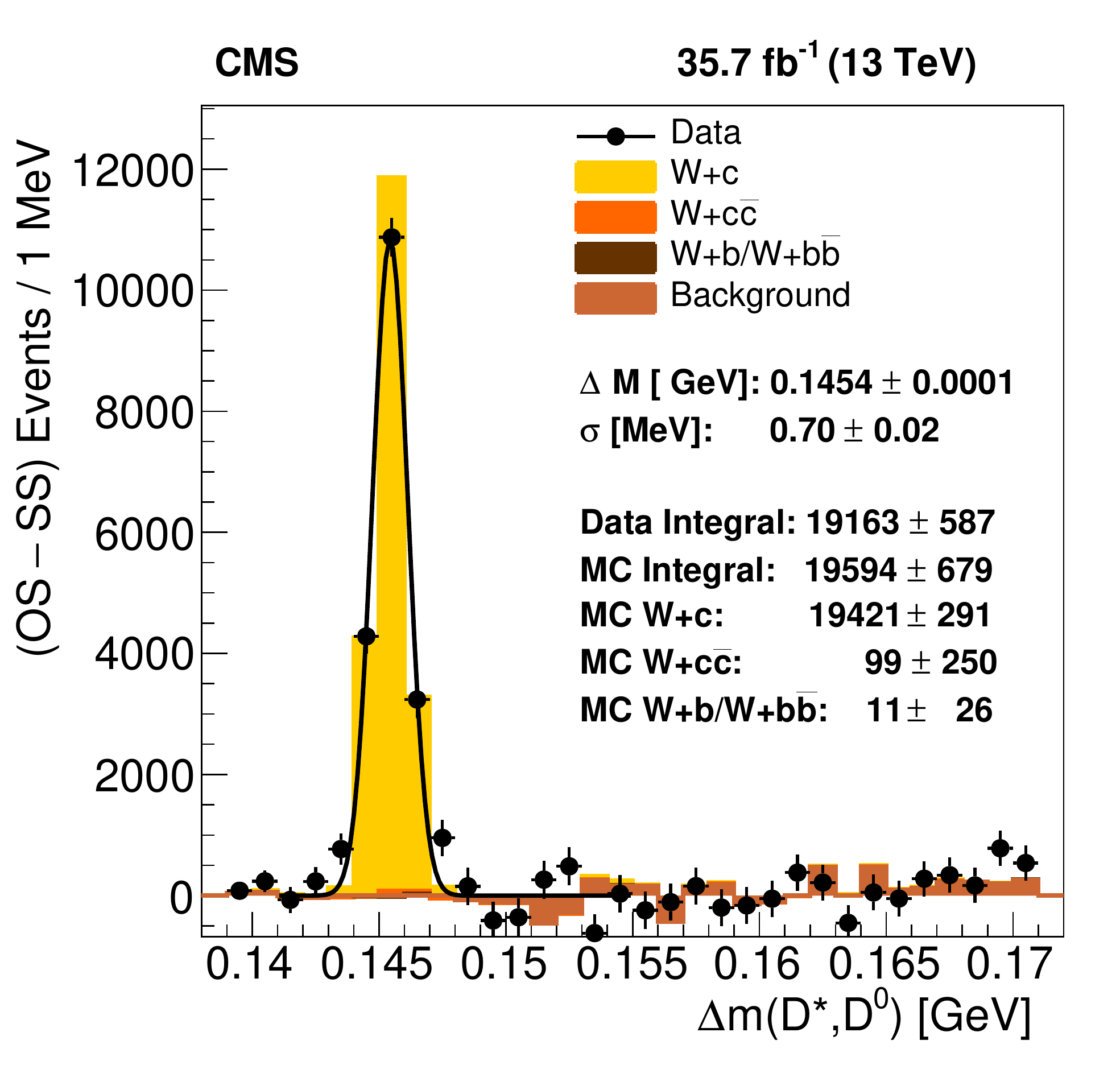}
	\caption{ Distributions of the reconstructed invariant mass of $\Kaonmp \Pgppm$ candidates (\cmsLeft) in the range $\abs{\deltam-0.1454}<0.001\GeV$, and the reconstructed mass difference \deltam (\cmsRight). The SS combinations are subtracted. The data (filled circles) are compared to MC simulation with contributions from different processes shown as histograms of different shades.	
	}
	\label{Dzero_DeltaM}
\end{figure}

{\tolerance=800
The number of \wc events corresponds to the number of \Dstar mesons after the subtraction of light-flavor and gluon splitting backgrounds.
The invariant mass of $\Kaonmp \Pgppm$ candidates, which are selected in a \deltam window of $\pm$1\MeV, is shown in Fig.~\ref{Dzero_DeltaM}, along with the observed reconstructed mass difference \deltam.
A clear $\PDz$ peak at the expected mass and a clear \deltam  peak around the expected value of $145.4257 \pm 0.0017\MeV$ \cite{PDG2018} are observed.
The remaining background is negligible, and the number of \Dstar mesons is determined by counting the number of candidates in a window of $144 < \deltam < 147\MeV$.
Alternately, two different functions are fit to the distributions, and their integral over the same mass window is used to estimate the systematic uncertainties associated with the method chosen.
\par}

\section{Measurement of the fiducial \wc cross section}
\label{CrossSection}

The fiducial cross section is measured in a kinematic region defined by requirements on the transverse momentum and the pseudorapidity of the muon and the transverse momentum of the charm quark.
The simulated signal is used to extrapolate from the fiducial region of the \Dstar meson to the fiducial region of the charm quark.
Since the \Dstar kinematics is integrated over at the generator level, the only kinematic constraint on the corresponding
charm quark arises from the requirement on the transverse momentum of \Dstar meson.
The correlation of the kinematics of charm quarks and \Dstar mesons is investigated using simulation, and the requirement of $\ptDstar > 5\GeV$  translates into $\ptCharm > 5\GeV$.
The distributions of $\etaMuon$ and $\ptCharm$ in the simulation are shown to reproduce very well the fixed order prediction at NLO obtained, using {\MCFM 6.8}~\cite{Campbell:1999ah,Campbell:2010ff,Campbell:2012uf} calculation.
The kinematic range of the measured fiducial cross section corresponds to $\ptMuon > 26\GeV$, $\etaMuon < 2.4$, and $\ptCharm > 5\GeV$.

The fiducial \wc cross section is determined as:
\begin{equation}
	\sigmawc =  \frac{\Nsel \, \mathcal{S}}{\large \intLumi \, \mathcal{B} \, \mathcal{C} },
	\label{Inc_CrossSection_Formula}
\end{equation}
where $\Nsel$ is the number of selected $\mbox{OS} - \mbox{SS}$ events in the \deltam distribution and $\mathcal{S}$ is the signal fraction.
The latter is defined as the ratio of the number of reconstructed \wDstar candidates originating from \wc to the number of all reconstructed \Dstar.
It is determined from the MC simulation, includes the background contributions, and varies between 0.95 and 0.99.
The integrated luminosity is denoted by $\intLumi$.
The combined branching fraction $\mathcal{B}$ for the channels under study is a product of $\mathcal{B}(\cPqc \to \Dstar) = 0.2429$ $\pm$ 0.0049~\cite{Lisovyi:2015uqa} and $\mathcal{B}(\PDstpm \to \Kaonmp + \Pgppm + \PiSlowpm)$ = 0.0266 $\pm$ 0.0003 \cite{PDG2018}.
The correction factor $\mathcal{C}$ accounts for the acceptance and efficiency of the detector.
The latter is determined using the MC simulation and is defined as the ratio of the number of reconstructed \wDstar candidates to the number of generated \wDstar originating from \wc events that fulfill the fiducial requirements.
In the measurement of the \wpluscbar (\wpDstar) and \wminusc (\wmDstar) cross sections, the factor $\mathcal{C}$ is determined separately for different charge combinations.

The measurement of the \wc cross section relies to a large extent on the MC simulation and requires extrapolation to unmeasured phase space.
To reduce the extrapolation and the corresponding uncertainty, the cross section for \wDstar production is also determined in the fiducial phase space of the detector-level measurement, $\ptMuon > 26\GeV$, $\etaMuon < 2.4$, $\etaDstar < 2.4$  and $\ptDstar > 5\GeV$, in a similar way by modifying Eq.~(\ref{Inc_CrossSection_Formula}) as follows:
only the branching fraction $\mathcal{B} = \mathcal{B}(\PDstpm \to \Kaonmp+\Pgppm+\PiSlowpm)$ is considered and the factor $\mathcal{C}$ is defined as the ratio between the numbers of reconstructed and of generated \wDstar candidates in the fiducial phase space after $\mbox{OS} - \mbox{SS}$ subtraction.

\begin{figure}[hb!]
	\centering
	\includegraphics[width=0.48\textwidth]{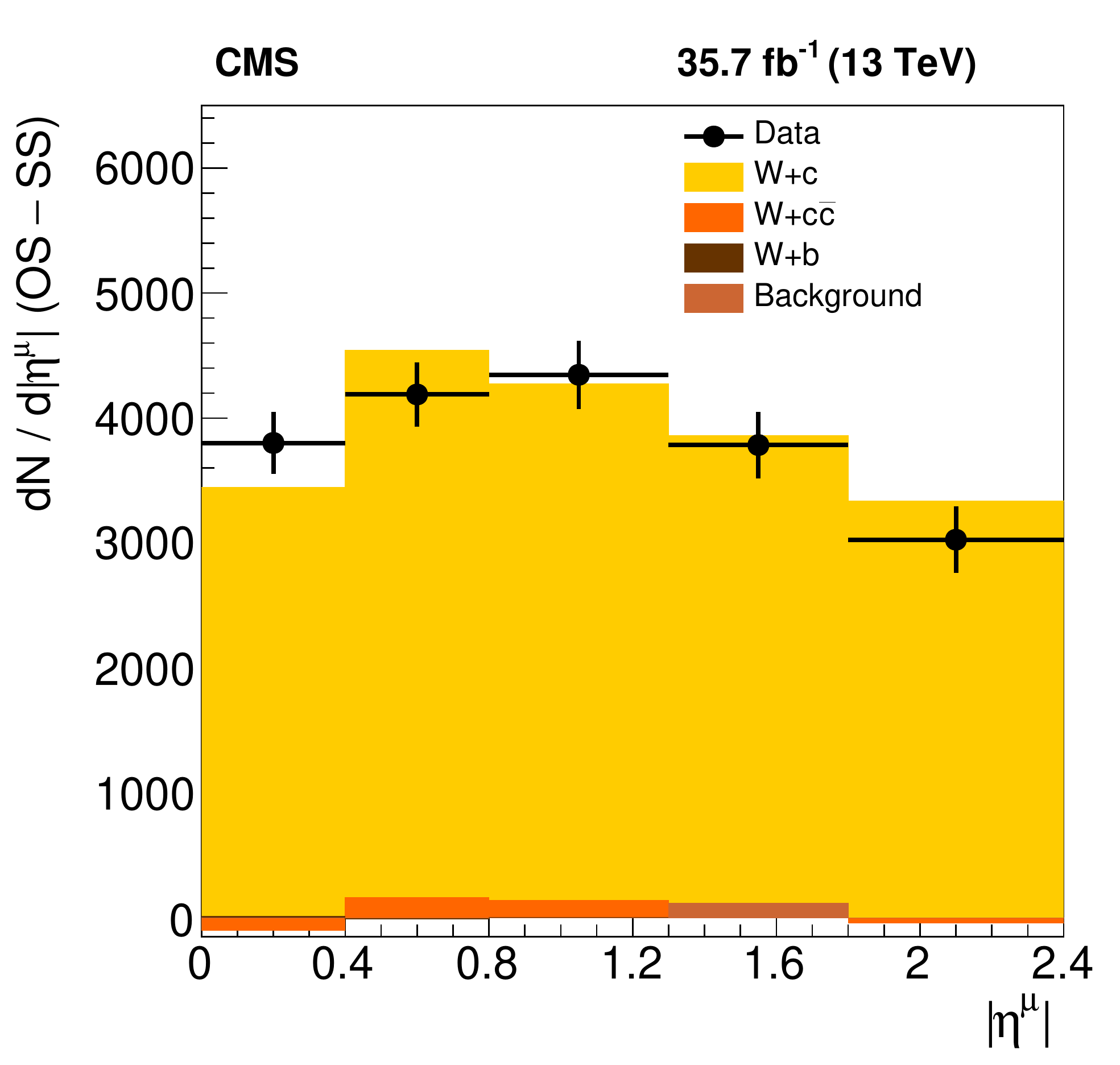}
	\caption{Number of events after $\mbox{OS} - \mbox{SS}$ subtraction for data (filled circles) and MC simulation (filled histograms) as a function of $\etaMuon$.}
	\label{OS_SS_Differential}	
\end{figure}

The cross sections are determined inclusively and also in five bins of the absolute pseudorapidity $\etaMuon$ of the muon originating from the $\PW$ boson decay.
The number of signal ($\mbox{OS} - \mbox{SS}$) events in each range of $\etaMuon$ is shown in Fig.~\ref{OS_SS_Differential}.
Good agreement between the data and MC simulation within the statistical uncertainties is observed.

\subsection{Systematic uncertainties}
The efficiencies and the assumptions relevant for the measurement are varied within their uncertainties to estimate the systematic uncertainty in the cross section measurement.
The resulting shift of the cross section with respect to the central result is taken as the corresponding uncertainty contribution.
The various sources of the systematic uncertainties in the \wc production cross section are listed in Table~\ref{Diff_Systematics} for both the inclusive and the differential measurements.

\begin{itemize}

\item Uncertainties associated with the integrated luminosity measurement are estimated as 2.5\%~\cite{CMS-PAS-LUM-17-001}.

\item The uncertainty in the tracking efficiency is 2.3\% for the 2016 data.
It is determined using the same method described in Ref.~\cite{CMS-PAS-TRK-10-002}, which exploits the ratio of branching fractions between the four-body and two-body decays of the neutral charm meson to all-charged final states.

\item The uncertainty in the branching fraction of the $\cPqc \to \Dstar$ is 2.4\%~\cite{Lisovyi:2015uqa}.

\item The muon systematic uncertainties are 1\% each for for the muon identification and isolation, and 0.5\% for the trigger and tracking corrections.
These are added in quadrature and the resulting uncertainty for muons is 1.2\%, which is referred to as the 'muon uncertainty'.

\item The uncertainty in the determination of $\Nsel$ is estimated from the difference in using a Gaussian or Crystal Ball fit~\cite{CrystalBallRef}.
The largest value of this uncertainty determined differentially, 1.5\%, is considered for all.

\item Uncertainties in the modeling of kinematic observables of the generated \Dstar meson are estimated by reweighting the simulated $\ptDstar$ and $\etaDstar$ distributions to the shape observed in data. The respective uncertainty in the inclusive cross section measurement is 0.5\%.
Due to statistical limitations, this uncertainty is determined inclusively in \etaMuon.

\item The uncertainty in the difference of the normalization of the \deltam distributions for $\Kaonmp \Pgppm \Pgppm$ and $\PKpm \Pgppm \Pgp^\mp$ combinations ('background normalization') is 0.5\%.

\item Uncertainties in the measured $\ptvecmiss$ are estimated in Ref.~\cite{CMS-PAS-JME-16-004} and result in an overall uncertainty of 0.9\% for this analysis.

\item Uncertainties due to the modeling of pileup are estimated by varying the total inelastic cross section used in the simulation of pileup events by 5\%.
The corresponding uncertainty in the \wc cross section is 2\%.

{\tolerance=800
\item The uncertainty related to the requirement of a valid secondary vertex, fitted from the tracks associated with a $\PDz$ candidate, is determined by calculating the \Dstar reconstruction efficiency in data and MC simulation for events with and without applying this selection criterion.
The number of reconstructed \Dstar candidates after the SS event subtraction is compared for events with or without a valid secondary vertex along with the proximity requirement ($\Delta_{xy} < 0.1\unit{cm}$, $\Delta_z < 0.1\unit{cm}$).
The difference in efficiency between data and MC simulation is calculated and an uncertainty in the inclusive cross section of $-1.1$\% is obtained.
Since this variation is not symmetric, the uncertainty is one-sided.
\par}
\item The PDF uncertainties are determined according to the prescription of the PDF group \cite{Ball:2014uwa}.
These are added in quadrature to the uncertainty related to the variation of \asmz in the PDF, resulting in an uncertainty of 1.2\% in the inclusive cross section.

\item The uncertainty associated with the fragmentation of the $\cPqc$ quark into a \Dstar meson is determined through variations of the function describing the fragmentation parameter $z = \pt^{\PD^*}/\ptCharm$.
The investigation of this uncertainty is inspired by a dedicated measurement of the $\cPqc \to \Dstar$ fragmentation function in electron-proton collisions~\cite{Aaron:2008ac},
in which the fragmentation parameters in various phenomenological models were determined with an uncertainty of 10\%.
In the \PYTHIA MC event generator, the fragmentation is described by the phenomenological Bowler--Lund function~\cite{Bowler:1981sb,Andersson:1983ia}, in the form
\begin{equation*}
f(z) = \frac{1}{z^{r_c \, b \, m_q^2}} (1-z)^a \exp(-b \, m_\perp^2 / z) \, c,
\end{equation*}
with   $m_\perp = \sqrt{\smash[b]{m_{\PD^*}^2 + p_{\mathrm{T\PD^*}}^2}},$
controlled by the two parameters $a$ and $b$. In the case of charm quarks, $r_c$ = 1 and  $m_q = 1.5\GeV$ are the \PYTHIA standard settings in the CUETP8M1 tune, whereas the value of $m_\perp$ is related to the average transverse momentum of generated \Dstar in the MC sample.
The parameters $a$, $b$ and $c$ are determined in a fit to the simulated distribution of $f(z)$, where $c$ is needed for the normalization.
Since the presence of a jet is not required in the analysis, the charm quark transverse momentum is approximated by summing up the transverse momenta of tracks in a cone of $\Delta R \leq$ 0.4 around the axis of the \Dstar candidate.
The free parameters are determined as $a = 1.827 \pm 0.016$ and $b = 0.00837 \pm 0.00005\GeV^{-2}$.
To estimate the uncertainty, the parameters $a$ and $b$ are varied within $\pm$10\% around their central values, following the precision achieved for the fragmentation parameters in \cite{Aaron:2008ac}. An additional constraint on the upper boundary on the $a$ parameter in \PYTHIA is consistent with this 10\% variation.
The resulting uncertainty in the cross section is 3.9\%.
\end{itemize}

\begin{table*}[ht]
	\topcaption{Systematic uncertainties [\%] in the inclusive and differential \wc cross section measurement in the fiducial region of the analysis. The total uncertainty corresponds to the sum of the individual contributions in quadrature.
	 The contributions listed in the top part of the table cancel in the ratio $\swpluscbar / \swminusc$.	 }
\centering
	\cmsTable{
	\begin{tabular}{l c c c c c c}
	\hline	

	Pseudorapidity $[\etaMuon]$& $[0, 2.4]$& $[0, 0.4]$ & $[0.4, 0.8]$ & $[0.8, 1.3]$ & $[1.3, 1.8]$ & $[1.8, 2.4]$ \\\hline
	Luminosity 			&	$\pm$2.5	& $\pm$2.5 & $\pm$2.5 & $\pm$2.5 & $\pm$2.5 & 	$\pm$2.5	\\				
	Tracking			&	$\pm$2.3	& $\pm$2.3 & $\pm$2.3 & $\pm$2.3 & $\pm$2.3 & 	$\pm$2.3	\\					
	Branching 			&	$\pm$2.4	& $\pm$2.4 & $\pm$2.4 & $\pm$2.4 & $\pm$2.4 & 	$\pm$2.4	\\				
	Muons				&	$\pm$1.2	& $\pm$1.2 & $\pm$1.2 & $\pm$1.2 & $\pm$1.2 & 	$\pm$1.2	\\
    $\Nsel$ determination &	$\pm$1.5	& $\pm$1.5 & $\pm$1.5 & $\pm$1.5 & $\pm$1.5 & 	$\pm$1.5	\\
    $\Dstar$  &	 $\pm$0.5	& $\pm$0.5  & $\pm$0.5	& $\pm$0.5	& $\pm$0.5	&	$\pm$0.5	\\			
    kinematics	&	&&&&&\\
	&	&&&&&\\
	
	Background &&&&&&\\
	normalization		&	$\pm$ 0.5	& $+$0.9/$-$0.8 & $+$1.9/$-$0.8 & $+$1.4/$-$0.5 & $+$0.8/$-$1.0   & 0.0/$-$0.6		\\				
	$\ptvecmiss$	 	&	$+$0.7/$-$0.9	& $+$0.4/$-$1.2 & $+$1.3/$-$0.3 & $+$1.1/$-$1.0 & 0.0/$-$2.6 	      & 0.0/$+$1.5		\\				
	Pileup 		 		&	$+$2.0/$-$1.9	& $+$0.4/$-$0.5 & $+$2.9/$-$3.0 & $+$2.0/$-$1.9 & $+$4.6/$-$5.1   & $+$2.7/$-$2.6	\\
	Secondary vertex	&	$-$1.1		& $+$1.3 		& $-$1.2 	    & $-$1.5		& $-$2.7 	      & $-$2.5		\\ 			
	PDF 			 	&	$\pm$1.2	& $\pm$1.3  & $\pm$0.9  & $\pm$1.4  & $\pm$1.5    & $\pm$1.7	\\					
	Fragmentation		&	$+$3.9/$-$3.2	& $+$3.4/$-$1.8 & $+$7.4/$-$5.2 & $+$3.3/$-$3.0 & $+$2.2/$-$1.2   & $+$7.4/$-$5.7	\\
	MC statistics		&	$+$3.6/$-$3.3 	& $+$8.8/$-$7.5 & $+$9.0/$-$11.9 & $+$7.9/$-$6.8 & $+$9.8/$-$14.1 & $+$10.1/$-$8.5 \\[\cmsTabSkip]
		&	&&&&&\\
	Total 			 	&	$+$7.5/$-$7.0	& $+$10.7/$-$9.3 & $+$13.2/$-$14.2 & $+$10.1/$-$9.3 & $+$12.7/$-$16.2 & $+$13.8/$-$12.1	\\ \hline	
	\end{tabular}
	}
	\label{Diff_Systematics}
\end{table*}

\subsection {Cross section results}
The numbers of signal events and the inclusive fiducial cross sections with their uncertainties are listed in Table~\ref{Inclusive_Cross_Section_table} together with the ratio of $\swpluscbar / \swminusc$.
For the differential measurement of the \wc cross section, the numbers of signal events are summarized in Table~\ref{Differential_Cross_Section_Table} together with the corrections $\mathcal{C}$ derived using MC simulations in each $\etaMuon$ bin.
The results are presented for $\dsigmawc/\detaMuon$, as well as for $\dswpluscbar/\detaMuon$ and for $\dswminusc/\detaMuon$.

\begin{table*}[th]
	\centering
	\topcaption{Inclusive cross sections of \wc and \wDstar production in the fiducial range of the analysis. The correction factor $\mathcal{C}$ accounts for the acceptance and efficiency of the detector. }
	\begin{tabular}{l r r r}
	\hline			& \multicolumn{1}{c}{\wc}						&  \multicolumn{1}{c}{\wpluscbar}	&	\multicolumn{1}{c}{\wminusc}		\\\hline
	$\Nsel $		&	19210 $\pm$	587\stat	   					&	9674 $\pm$	401\stat			&	9546 $\pm$ 367\stat					\\
	$\mathcal{C}$	&	0.0811 $\pm$ 0.003\stat	    				& 	0.0832 $\pm$ 0.004\stat			&	0.0794 $\pm$ 0.004\stat				\\[\cmsTabSkip] 	
	$\sigma$ [pb]   &  1026 $\pm$ 31\stat $\substack{+76\\-72}$\syst  & 504  $\pm$ 21\stat  $\pm$42\syst     & 521 $\pm$ 20\stat  $\substack{+42\\-40}$\syst \\[\cmsTabSkip]
	\multicolumn{2}{l}{$\frac{\swpluscbar}{\swminusc}$}&  \multicolumn{2}{c}{0.968 $\pm$ 0.055\stat $\substack{+0.015\\-0.028}$} \\ \hline 	
	&&&\\
	
	\hline			& \multicolumn{1}{c}{\wDstar}				&  \multicolumn{1}{c}{\wpDstar}		&	\multicolumn{1}{c}{\wmDstar}		\\ \hline
	$\Nsel $		&	19210 $\pm$	587\stat	   					&	9674 $\pm$	401\stat			&	9546 $\pm$ 367\stat					\\
	$\mathcal{C}$	&	0.107 $\pm$ 0.004\stat	    				& 	0.113  $\pm$ 0.006\stat		&	0.101 $\pm$ 0.004\stat				\\[\cmsTabSkip]
	$\sigma$ [pb]   &	190  $\pm$  6\stat $\substack{+12\\-13}$\syst & 	90 $\pm$ 4\stat $\substack{+7\\-8}$\syst & 	99  $\pm$ 3\stat 	$\pm$7\syst 	\\[\cmsTabSkip]
	\multicolumn{2}{l}{$\frac{\swpDstar}{\swmDstar}$}&  \multicolumn{2}{c}{0.909 $\pm$ 0.051\stat $\substack{+0.014\\-0.028}$} \\ \hline
	\end{tabular}
	\label{Inclusive_Cross_Section_table}
\end{table*}

\begin{table*}[ht]
	\topcaption{Number of signal events, correction factors $\mathcal{C}$, accounting for the acceptance and efficiency of the detector and the differential cross sections in each $\etaMuon$ range for \wc (upper), \wpluscbar (middle) and \wminusc (lower).}
\centering
	\vspace{0.2cm}
	\large $\mathbf \PW + \cPqc$\\ \normalsize
	\vspace{0.2cm}

	\begin{tabular}{l c c c }
	\hline\\[-2.2ex]
	$\etaMuonMinMax$	& $\Nsel $ & $\mathcal{C}$ & $\frac{\dsigmawc}{\detaMuon}$ [pb]	\\\hline\\[-2.2ex]
	$[0, 0.4]$			&	3795 $\pm$ 248\stat	&	0.072 $\pm$ 0.006\stat	&	569 $\pm$ 37\stat $\substack{+61\\-53}$ 	\\\\[-2.2ex]
	$[0.4, 0.8]$		&	4201 $\pm$ 256\stat	&	0.096 $\pm$ 0.006\stat	&	467 $\pm$ 28\stat $\substack{+61\\-66}$  \\\\[-2.2ex]	
	$[0.8, 1.3]$		&	4334 $\pm$ 274\stat	&	0.078 $\pm$ 0.006\stat	&	479 $\pm$ 30\stat $\substack{+49\\-45}$  \\\\[-2.2ex]
	$[1.3, 1.8]$		&	3823 $\pm$ 267\stat	&	0.083 $\pm$ 0.007\stat	&	395 $\pm$ 28\stat $\substack{+49\\-63}$  \\\\[-2.2ex]
	$[1.8, 2.4]$		&	3042 $\pm$ 266\stat	&	0.078 $\pm$ 0.007\stat	&	283 $\pm$ 25\stat $\substack{+39\\-34}$  \\\\[-2.2ex]
	\hline
	\end{tabular}
	
	\vspace{0.2cm}
	\large $\mathbf \PWp + \cPaqc$\\\normalsize
	\vspace{0.2cm}
	
	\begin{tabular}{ l c c c }
	\hline
	$\etaMuonMinMax$	& $\Nsel $ & $\mathcal{C}$ & $\frac{\dswpluscbar}{\detaMuon}$ [pb]	\\\hline\\[-2.2ex]
	$[0, 0.4]$			&	2109 $\pm$ 167\stat	&	0.073 $\pm$ 0.008\stat	&	313 $\pm$ 25\stat $\substack{+48\\-44}$ \\\\[-2.2ex]
	$[0.4, 0.8]$		&	2119 $\pm$ 172\stat	&	0.094 $\pm$ 0.010\stat	&	236 $\pm$ 19\stat $\substack{+37\\-41}$ \\\\[-2.2ex]	
	$[0.8, 1.3]$		&	2103 $\pm$ 186\stat	&	0.077 $\pm$ 0.008\stat	&	235 $\pm$ 21\stat $\substack{+33\\-27}$ \\\\[-2.2ex]
	$[1.3, 1.8]$		&	1840 $\pm$ 184\stat	&	0.093 $\pm$ 0.010\stat	&	162 $\pm$ 16\stat $\substack{+34\\-31}$ \\\\[-2.2ex]
	$[1.8, 2.4]$		&	1499 $\pm$ 186\stat	&	0.080 $\pm$ 0.011\stat	&	135 $\pm$ 17\stat $\substack{+24\\-26}$ \\\\[-2.2ex]\hline
	\end{tabular}
	
	\vspace{0.2cm}
	\large $\mathbf \PWm + \cPqc$\\\normalsize
	\vspace{0.2cm}
	
	\begin{tabular}{ l c c c }
	\hline
	$\etaMuonMinMax$	& $\Nsel $ & $\mathcal{C}$ & $\frac{\dswminusc}{\detaMuon}$ [pb]     \\\hline\\[-2.2ex]
	$[0, 0.4]$			&	1688 $\pm$ 158\stat	&	0.072 $\pm$ 0.008\stat	&	255 $\pm$ 23\stat $\substack{+35\\-42}$  \\\\[-2.2ex]
	$[0.4, 0.8]$		&	2084 $\pm$ 162\stat	&	0.097 $\pm$ 0.008\stat	&	231 $\pm$ 18\stat $\substack{+28\\-42}$  \\\\[-2.2ex]	
	$[0.8, 1.3]$		&	2234 $\pm$ 172\stat	&	0.079 $\pm$ 0.007\stat	&	244 $\pm$ 19\stat $\substack{+29\\-38}$  \\\\[-2.2ex]
	$[1.3, 1.8]$		&	1986 $\pm$ 166\stat	&	0.073 $\pm$ 0.008\stat	&	237 $\pm$ 20\stat $\substack{+33\\-37}$  \\\\[-2.2ex]
	$[1.8, 2.4]$		&	1544 $\pm$ 161\stat	&	0.075 $\pm$ 0.008\stat	&	149 $\pm$ 16\stat $\substack{+25\\-21}$  \\\\[-2.2ex]\hline
	\end{tabular}
	\label{Differential_Cross_Section_Table}	
\end{table*}

\begin{figure}
	\centering
	\includegraphics[width = 0.48\textwidth]{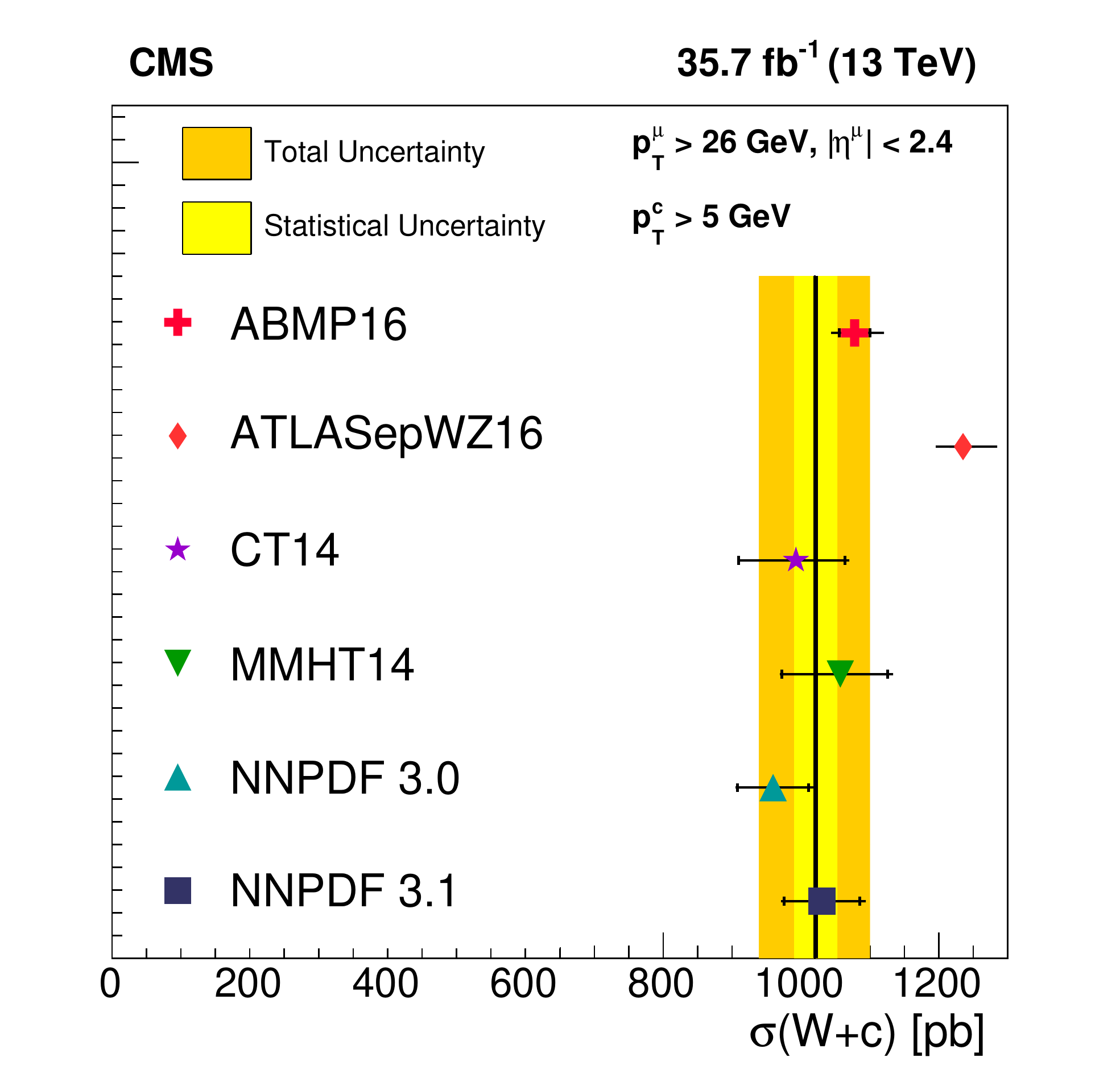}
	\includegraphics[width = 0.48\textwidth]{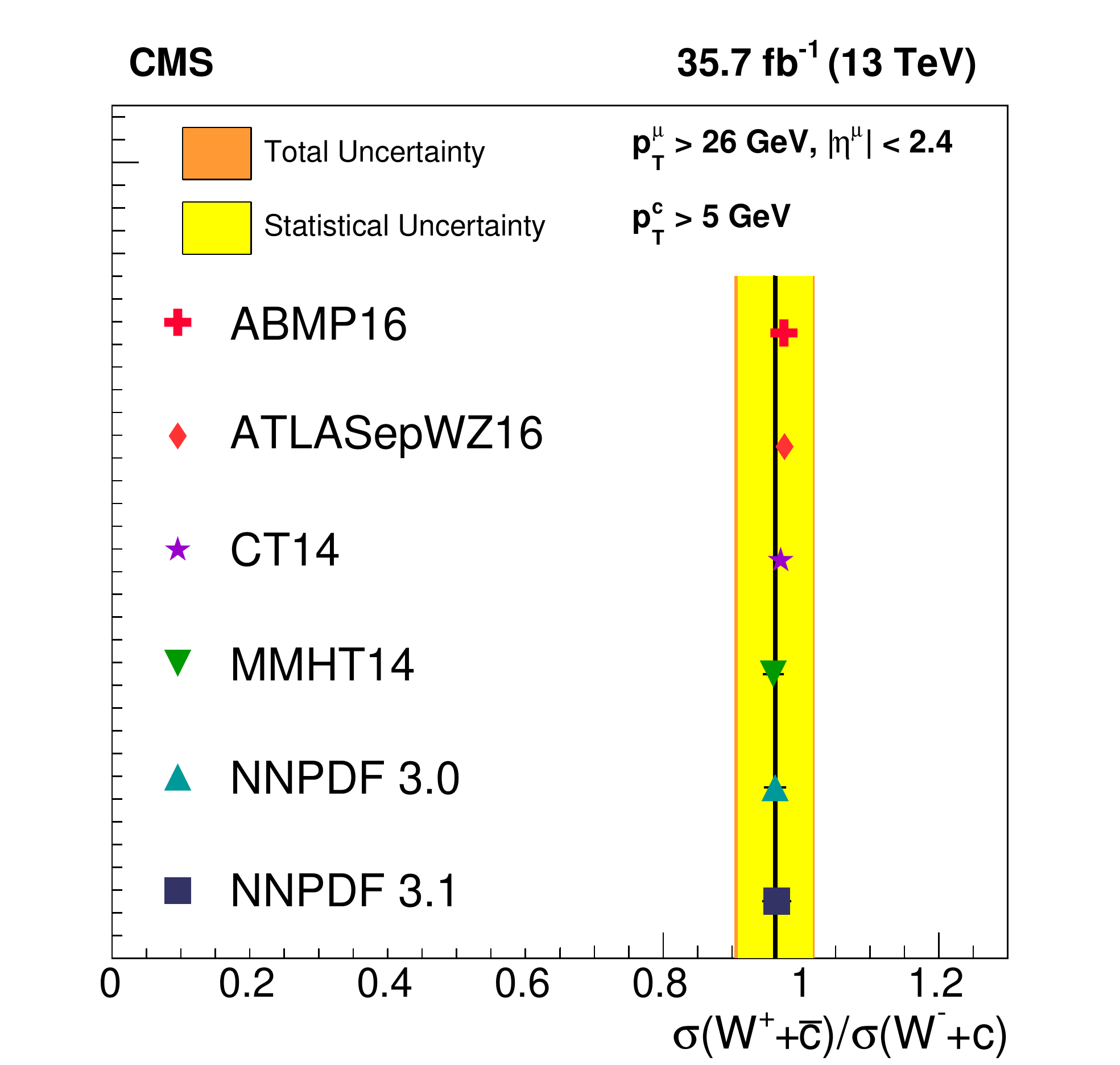}
	\caption{Inclusive fiducial cross section $\sigmawc$ and the cross section ratio $\swpluscbar/\swminusc$ at 13\TeV.
	The data are represented by a line with the statistical (total) uncertainty shown by a light (dark) shaded band.
	The measurements are compared to the NLO QCD prediction using several PDF sets, represented by symbols of different types.
	All used PDF sets are evaluated at NLO, except for ATLASepWZ16, which is obtained at NNLO.
	The error bars depict the total theoretical uncertainty, including the PDF and the scale variation uncertainty.}
	\label{Comparison_Inclusive_CrossSection}
\end{figure}

\begin{figure*}
	\centering
	\includegraphics[width = 0.48\textwidth]{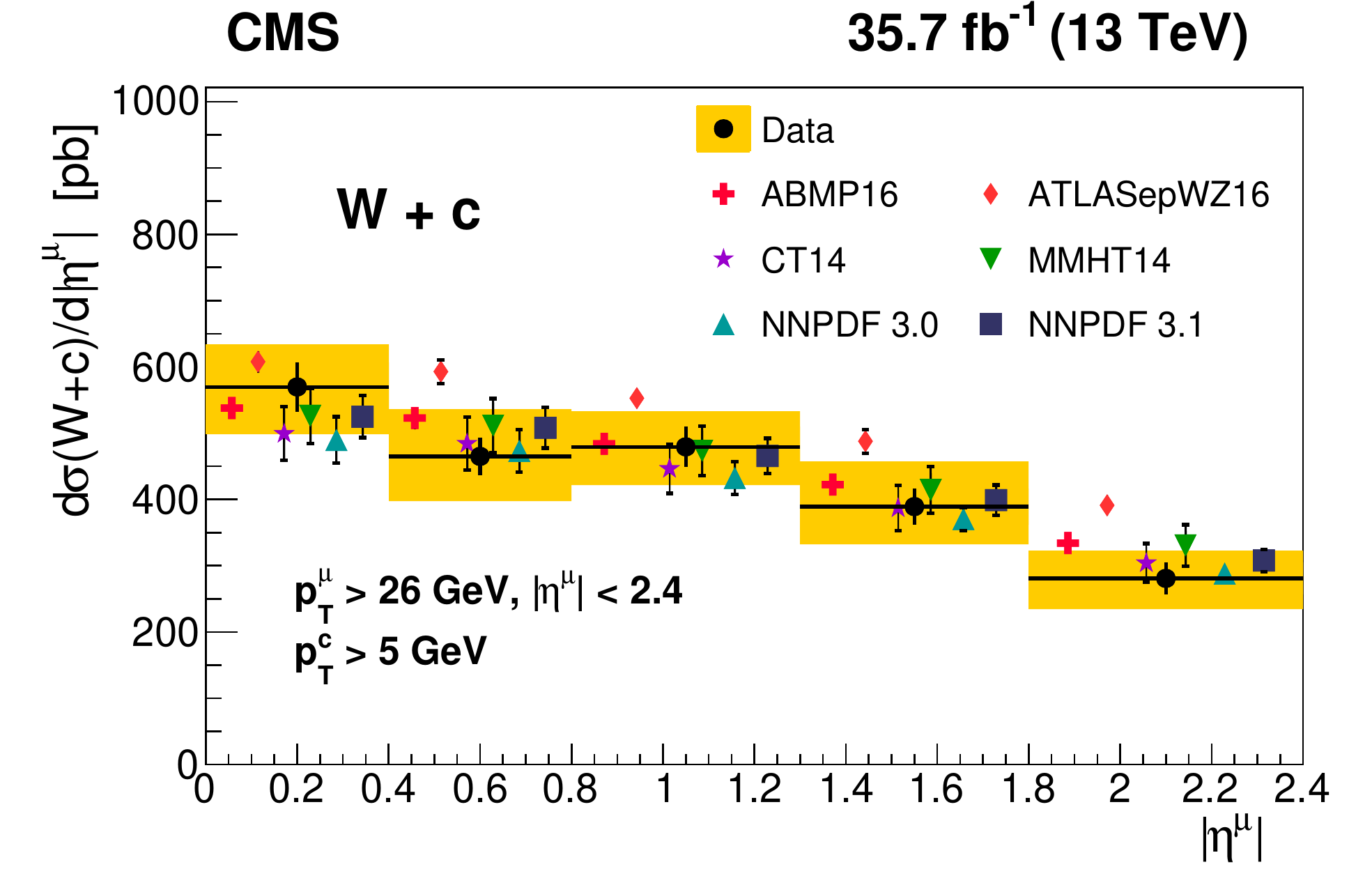}
	\includegraphics[width = 0.48\textwidth]{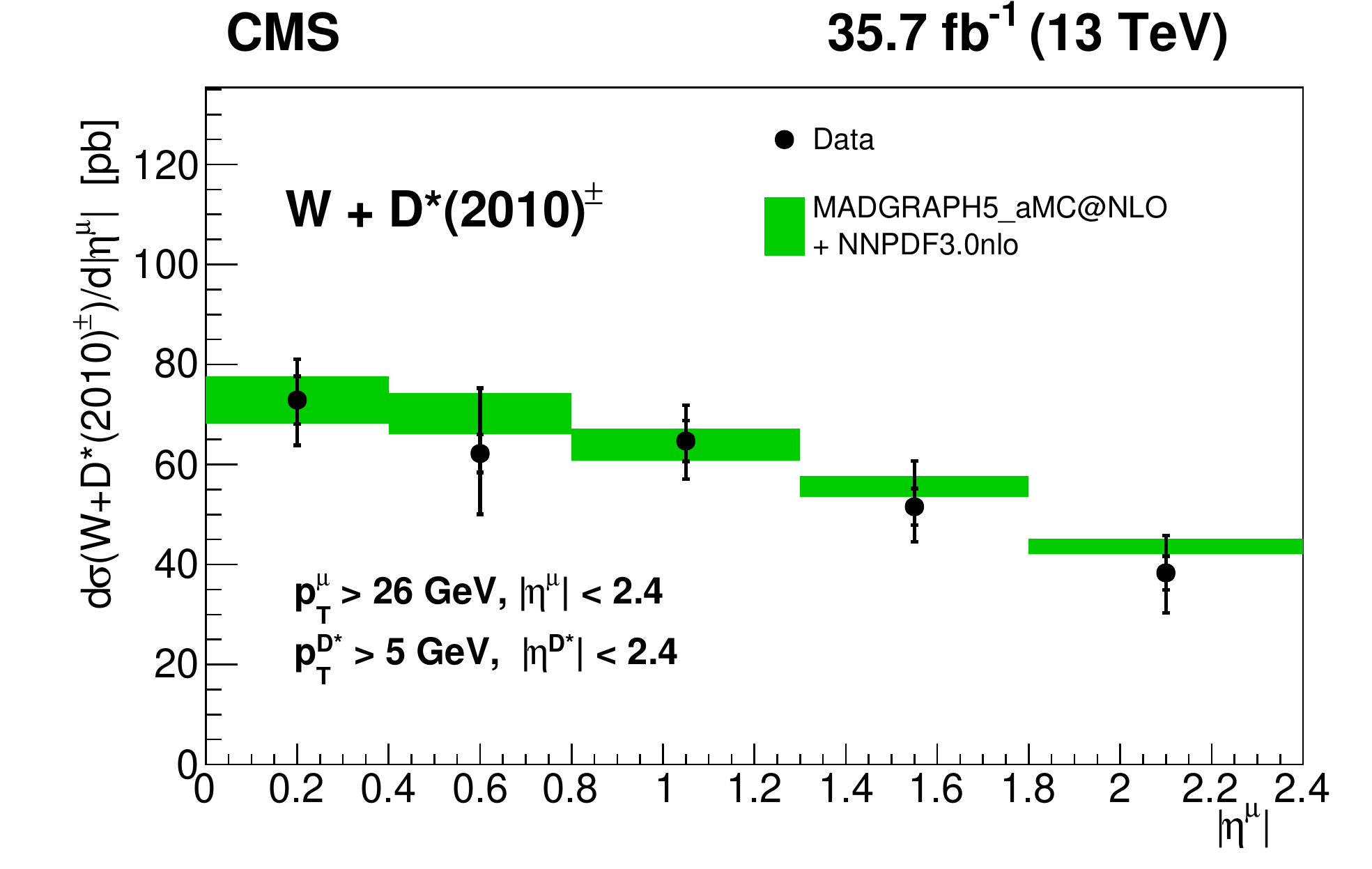}\\
	\includegraphics[width = 0.48\textwidth]{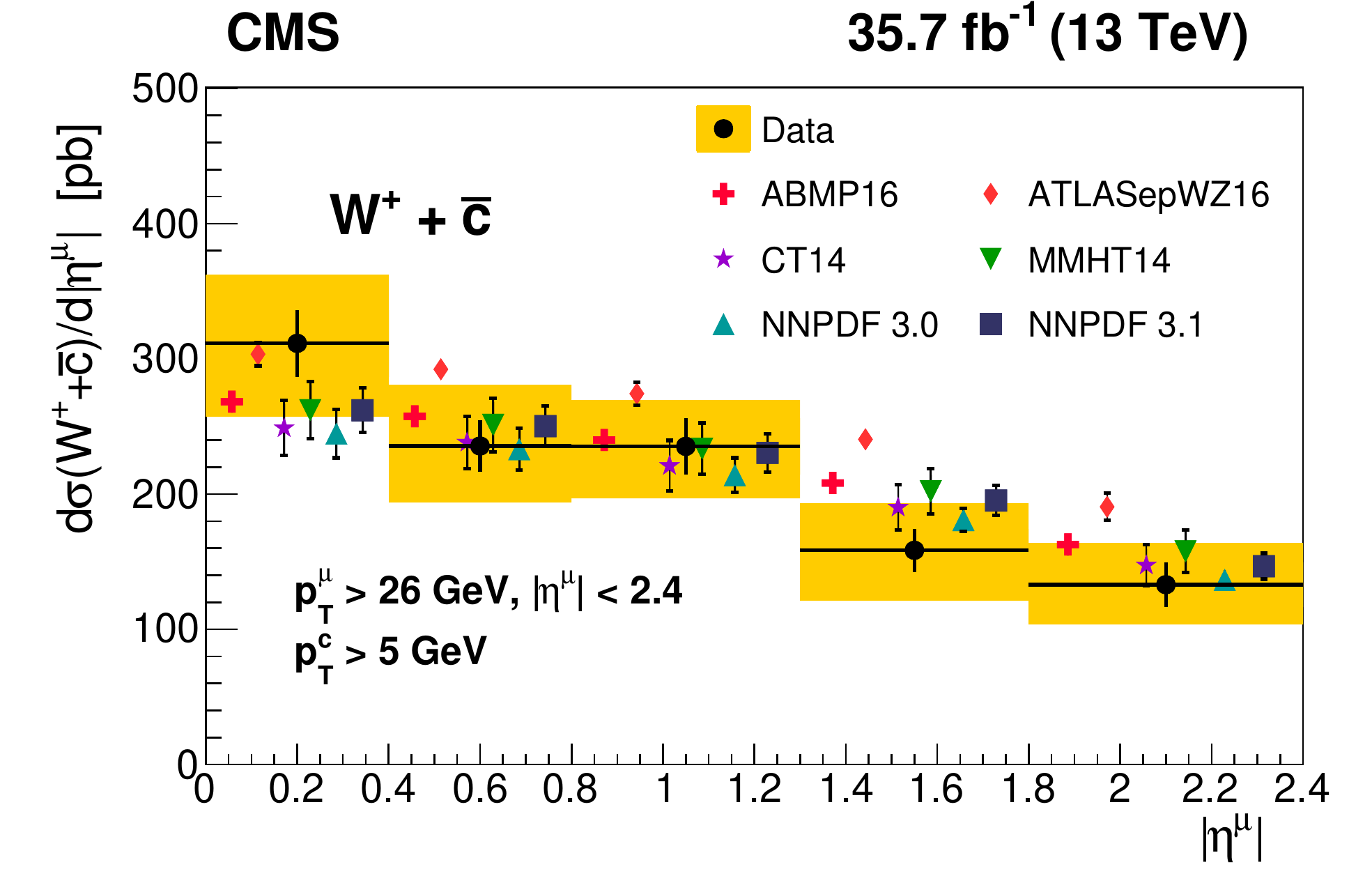}
	\includegraphics[width = 0.48\textwidth]{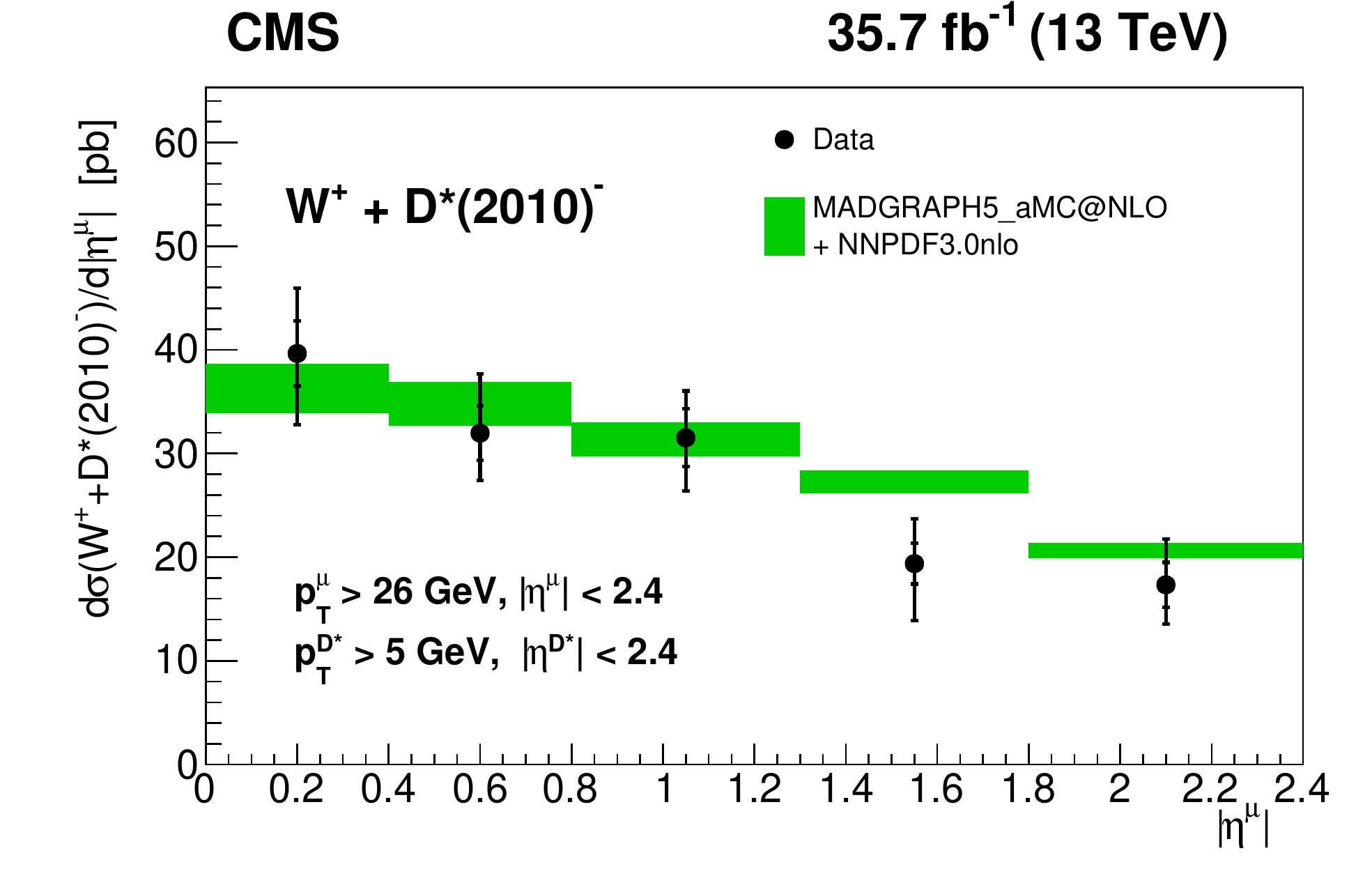}\\
	\includegraphics[width = 0.48\textwidth]{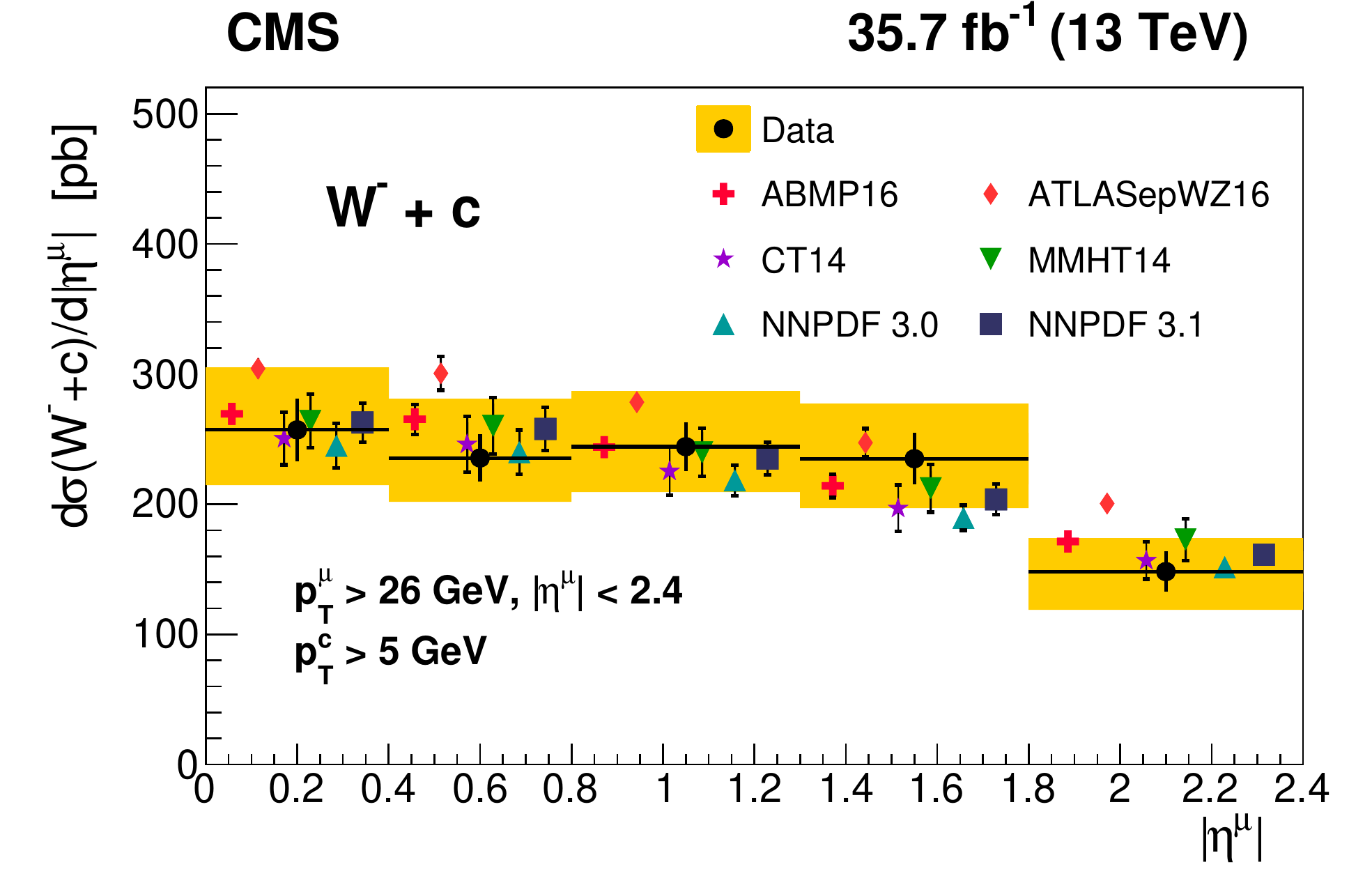}
	\includegraphics[width = 0.48\textwidth]{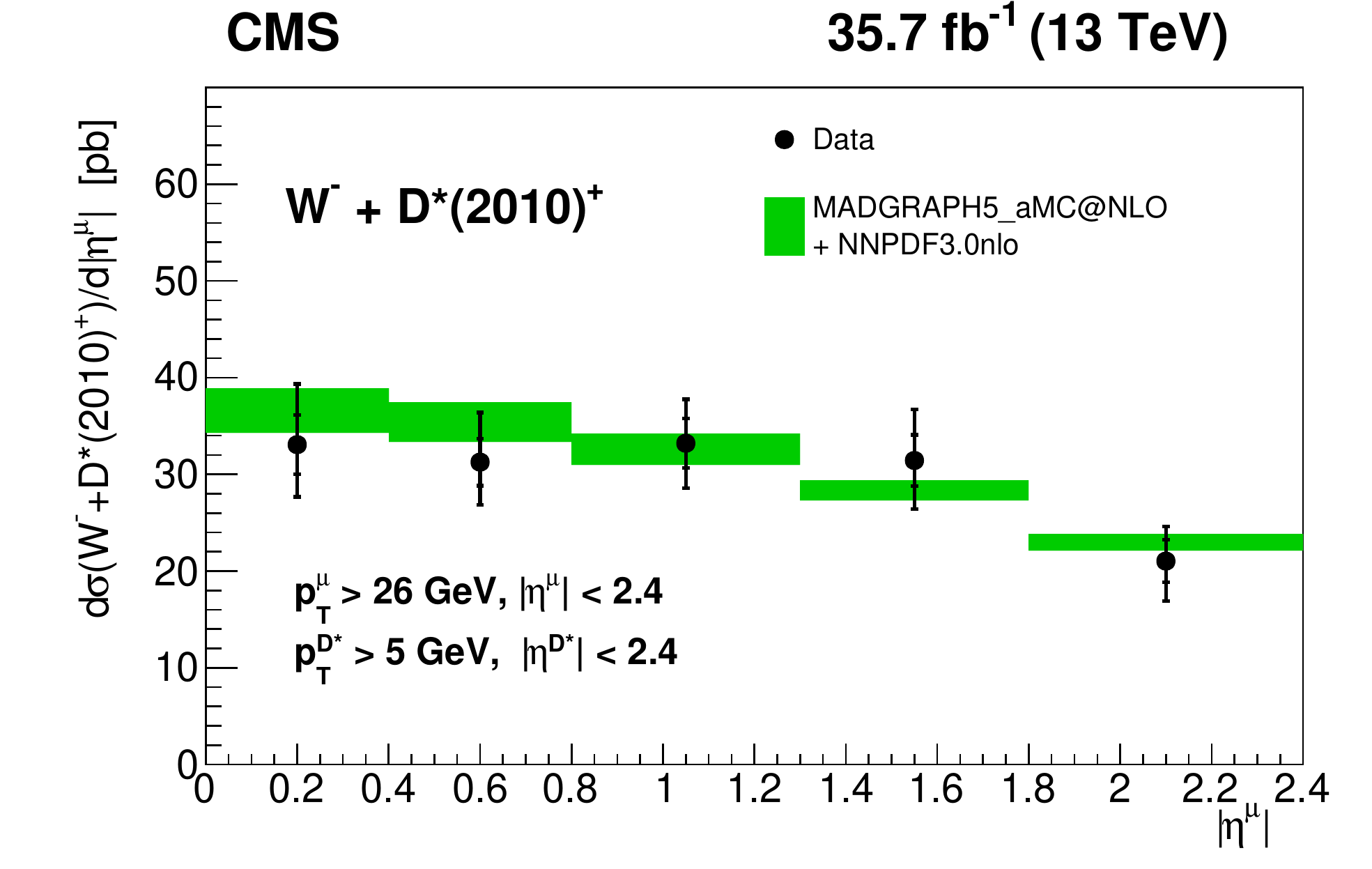}
	\caption{Left: differential cross sections of $\sigmawc$ production at 13\TeV measured as a function $\etaMuon$.
	The data are presented by filled circles with the statistical (total) uncertainties shown by vertical error bars (light shaded bands).
	The measurements are compared to the QCD predictions calculated with {\MCFM} at NLO using different PDF sets, presented by symbols of different style.
	All used PDF sets are evaluated at NLO, except for ATLASepWZ16, which is obtained at NNLO.
	The error bars represent theoretical uncertainties, which include PDF and scale variation uncertainty.
	Right: $\swDstar$ production differential cross sections presented as a function of $\etaMuon$.
	The data (filled circles) are shown with their total (outer error bars) and statistical (inner error bars) uncertainties and are compared to the predictions of the signal MC generated with \MGvATNLO and using NNPDF3.0nlo to describe the proton structure.
	PDF uncertainties and scale variations are accounted for and added in quadrature (shaded band).}
	\label{Comparison_Differential_CrossSection}
\end{figure*}

{\tolerance=1000
The measured inclusive and differential fiducial cross sections of \wc are compared to predictions at NLO ($\mathcal{O}(\alpha_s^2)$) that are obtained using {\MCFM 6.8}.
Similarly to the earlier analysis~\cite{Chatrchyan:2013mza}, the mass of the charm quark is chosen to be $m_{{c}} = 1.5\GeV$, and the factorization and the renormalization scales are set to the value of the $\PW$ boson mass.
The calculation is performed for $\ptMuon > 26\GeV$, $\etaMuon < 2.4$, and $\ptCharm > 5\GeV$.
In Fig.~\ref{Comparison_Inclusive_CrossSection}, the measurements of the inclusive \wc cross section and the charge ratio are compared to the NLO predictions calculated using the ABMP16nlo~\cite{Alekhin:2018pai}, ATLASepWZ16nnlo~\cite{Aaboud:2016btc}, CT14nlo~\cite{Dulat:2015mca}, MMHT14nlo~\cite{Harland-Lang:2014zoa}, NNPDF3.0nlo~\cite{Ball:2014uwa}, and NNPDF3.1nlo~\cite{Ball:2017nwa} PDF sets.
The values of the strong coupling constant \asmz are set to those used in the evaluation of a particular PDF.
The details of the experimental data, used for constraining the strange quark content of the proton in the global PDFs, are given in  Refs.~\cite{Alekhin:2017kpj,Aaboud:2016btc,Dulat:2015mca,Harland-Lang:2014zoa,Ball:2014uwa}.
In these references, the treatment of the sea quark distributions in different PDF sets is discussed, and the comparison of the PDFs is presented.
The ABMP16nlo PDF includes the most recent data on charm quark production in charged-current neutrino-nucleon DIS collected by the NOMAD and CHORUS experiments in order to improve the constraints on the strange quark distribution and to perform a detailed study of the isospin asymmetry of the light quarks in the proton sea~\cite{Alekhin:2015cza}.
Despite differences in the data used in the individual global PDF fits, the strangeness suppression distributions in ABMP16nlo, NNPDF3.1nlo, CT14nlo and MMHT14nlo are in a good agreement among each other and disagree with the ATLASepWZ16nnlo result~\cite{Aaboud:2016btc}.
\par }
{\tolerance=1000
The predicted inclusive cross sections are summarized in Table~\ref{MCFM_Predictions}.
The PDF uncertainties are calculated using prescriptions from each PDF group.
For the ATLASepWZ16nnlo~PDFs no respective NLO set is available and only Hessian uncertainties are considered in this paper.
For other PDFs, the variation of $\alpha_s(m_Z)$ is taken into account as well.
The uncertainties due to missing higher-order corrections are estimated by varying $\renormScale$ and $\factScale$ simultaneously by a factor of 2 up and down, and the resulting variation of the cross section is referred to as the scale uncertainty, $\Delta \PGm$.
Good agreement between NLO predictions and the measurements is observed, except for the prediction using ATLASepWZ16nnlo.
For the cross section ratio \swpluscbar/\swminusc, all theoretical predictions are in good agreement with the measured value.
In Table~\ref{MCFM_Differential_CrossSection}, the theoretical predictions for $\dsigmawc/\detaMuon$ using different PDF sets are summarized.
In Fig.~\ref{Comparison_Differential_CrossSection}, the measurements of differential \wc and \wDstar cross sections are compared with the {\MCFM} NLO calculations and with the signal MC prediction, respectively.
Good agreement between the measured \wc cross section and NLO calculations is observed except for the prediction using the ATLASepWZ16nnlo PDF set.
The signal MC prediction using NNPDF3.0nlo is presented with the PDF and $\alpha_s$ uncertainties and accounts for simultaneous variations of $\renormScale$ and $\factScale$ in the matrix element by a factor of 2.
The \wDstar cross section is described well by the simulation.
\par }

\begin{table*}
	\topcaption{The NLO predictions for \sigmawc, obtained with {\MCFM}~\cite{Campbell:1999ah,Campbell:2010ff,Campbell:2012uf}. The uncertainties account for PDF and scale variations.}
\centering
	\begin{tabular}{lcccc}
	\hline
					&	\sigmawc [pb]	& $\Delta$PDF [\%] & $\Delta \PGm$ [\%]		&	\swpluscbar/\swminusc 	\\ \hline\\[-2.2ex]
	ABMP16nlo 		& 	1077.9	&	$\pm$ 2.1 				&	$\substack{+3.4\\-2.4}$	& 0.975 $\substack{+0.002\\-0.002}$		\\\\[-2.2ex]
	ATLASepWZ16nnlo	& 	1235.1	&	$\substack{+1.4\\-1.6}$ & 	$\substack{+3.7\\-2.8}$	& 0.976 $\substack{+0.001\\-0.001}$		\\\\[-2.2ex]
	CT14nlo			& 	992.6	&	$\substack{+7.2\\-8.4}$ & 	$\substack{+3.1\\-2.1}$	& 0.970 $\substack{+0.005\\-0.007}$		\\\\[-2.2ex]
	MMHT14nlo		& 	1057.1	&	$\substack{+6.5\\-8.0}$ & 	$\substack{+3.2\\-2.2}$ & 0.960 $\substack{+0.023\\-0.033}$		\\\\[-2.2ex]
	NNPDF3.0nlo	& 	959.5	&	$\pm$ 5.4 				& 	$\substack{+2.8\\-1.9}$	& 0.962 $\substack{+0.034\\-0.034}$		\\\\[-2.2ex]
	NNPDF3.1nlo	& 	1030.2	&	$\pm$ 5.3				&	$\substack{+3.2\\-2.2}$	& 0.965 $\substack{+0.043\\-0.043}$		\\\\[-2.2ex]
	\hline		
	\end{tabular}
	\label{MCFM_Predictions}
\end{table*}

\begin{table*}[ht]
	\topcaption{Theoretical predictions for $\dsigmawc/\detaMuon$ calculated with {\MCFM} at NLO for different PDF sets. The relative uncertainties due to PDF and scale variations are shown.}
\centering
	\begin{tabular}{l ccc c ccc}
	\hline
						 	& \multicolumn{3}{c}{ABMP16nlo}					& 						&	\multicolumn{3}{c}{ATLASepWZ16nnlo}	\\[0.5ex]
	\hline\\[-2.2ex]								
	$\etaMuonMinMax$ 		& $\frac{\dsigmawc}{\detaMuon}$[pb] &	$\Delta$PDF[\%] & $\Delta \mu$ [\%]	& ~~ &$\frac{\dsigmawc}{\detaMuon}$[pb]	&	$\Delta$PDF[\%] & $\Delta \PGm$[\%]	 \\[0.5ex]
	\hline\\[-2.2ex]
	$[0, 0.4]$		&	537.8	&	$\pm$  2.2& 	$\substack{+3.7\\-1.9}$	& 	&	607.8	&	$\substack{+1.1\\-1.3}$ & 	$\substack{+4.2\\-2.4}$				\\[0.5ex]
	$[0.4, 0.8]$	&	522.8	&	$\pm$  2.1& 	$\substack{+3.1\\-2.3}$	&	&   592.9	&	$\substack{+1.1\\-1.3}$ & 	$\substack{+3.5\\-2.7}$ 			\\[0.5ex]
	$[0.8, 1.3]$	&	483.9	&	$\pm$  2.1& 	$\substack{+3.2\\-2.1}$	&	&   552.7	&	$\substack{+1.2\\-1.4}$ & 	$\substack{+3.6\\-2.5}$ 			\\[0.5ex]
	$[1.3, 1.8]$	&	422.4	&	$\pm$  2.0& 	$\substack{+3.4\\-2.9}$	&	&   487.8	&	$\substack{+1.4\\-1.6}$ & 	$\substack{+3.8\\-3.3}$ 			\\[0.5ex]
	$[1.8, 2.4]$	&	334.1	&	$\pm$  2.0& 	$\substack{+3.4\\-3.0}$	&	&   391.1	&	$\substack{+2.2\\-2.3}$ & 	$\substack{+3.6\\-3.3}$ 			\\[0.5ex]
	\hline
	\end{tabular}
	
	\vspace{0.5cm}
	\begin{tabular}{l ccc c ccc}
	\hline
						&	\multicolumn{3}{c}{CT14nlo}			&										& \multicolumn{3}{c}{MMHT14nlo}	\\
	\hline\\[-2.2ex]								
	$\etaMuonMinMax$	& 	$\frac{\dsigmawc}{\detaMuon}$[pb]	&	$\Delta$PDF[\%] & $\Delta \PGm$[\%]	& ~~ & $\frac{\dsigmawc}{\detaMuon}$[pb] &	$\Delta$PDF[\%] & $\Delta \PGm$[\%]  \\[0.5ex]
	\hline\\[-2.2ex]
	$[0, 0.4]$		&	499.3	&	$\substack{+7.0\\-8.0}$ &	$\substack{+3.4\\-1.7}$		&	&	526.0	&	$\substack{+7.0\\-7.7}$ &	$\substack{+3.6\\-1.8}$	 \\[0.5ex]
	$[0.4, 0.8]$	&	484.4	&	$\substack{+7.0\\-8.0}$ & 	$\substack{+2.9\\-2.1}$		&	&	511.2	&	$\substack{+6.8\\-7.7}$ & 	$\substack{+3.0\\-2.1}$  \\[0.5ex]
	$[0.8, 1.3]$	&	446.3	&	$\substack{+6.9\\-8.2}$ & 	$\substack{+2.9\\-1.8}$		&	&	473.4	&	$\substack{+6.4\\-7.7}$ & 	$\substack{+3.0\\-1.9}$  \\[0.5ex]
	$[1.3, 1.8]$	&	387.0	&	$\substack{+7.1\\-8.5}$ & 	$\substack{+3.1\\-2.6}$		&	&	414.4	&	$\substack{+6.0\\-8.0}$ & 	$\substack{+3.2\\-2.7}$	 \\[0.5ex]
	$[1.8, 2.4]$	&	304.1	&	$\substack{+7.8\\-9.3}$ & 	$\substack{+3.0\\-2.6}$		&	&	330.5	&	$\substack{+6.5\\-9.1}$ & 	$\substack{+3.2\\-2.7}$	 \\[0.5ex]
	\hline
	\end{tabular}
	
	\vspace{0.5cm}
	\begin{tabular}{l ccc c ccc}
	\hline
						&	\multicolumn{3}{c}{NNPDF3.0nlo}		&										& \multicolumn{3}{c}{NNPDF3.1nlo}	\\	
	\hline\\[-2.2ex]							
	$\etaMuonMinMax$	&$\frac{\dsigmawc}{\detaMuon}$[pb]	&	$\Delta$PDF[\%] & $\Delta \PGm$[\%]	& ~~ &$\frac{\dsigmawc}{\detaMuon}$[pb]	&	$\Delta$PDF[\%] & $\Delta \PGm$[\%] \\[0.5ex]
	\hline\\[-2.2ex]
	$[0, 0.4]$		&	489.8	&	$\pm$ 7.0 & 	$\substack{+3.2\\-1.5}$	&	&	524.8	&$\pm$	5.8 & 	$\substack{+3.6\\-1.8}$ \\[0.5ex]
	$[0.4, 0.8]$	&	473.2	&	$\pm$ 6.5 & 	$\substack{+2.7\\-1.8}$	&	&	508.1	&$\pm$	5.6 & 	$\substack{+3.0\\-2.2}$ \\[0.5ex]
	$[0.8, 1.3]$	&	432.4	&	$\pm$ 5.5 & 	$\substack{+2.6\\-1.5}$	&	&	465.6	&$\pm$	5.4 & 	$\substack{+3.0\\-1.9}$	\\[0.5ex]
	$[1.3, 1.8]$	&	370.4	&	$\pm$ 4.2 & 	$\substack{+2.7\\-2.3}$	&	&	399.0	&$\pm$	5.0 & 	$\substack{+3.1\\-2.7}$	\\[0.5ex]
	$[1.8, 2.4]$	&	288.1	&	$\pm$ 3.5 & 	$\substack{+2.7\\-2.3}$	&	&	307.9	&$\pm$	4.8 & 	$\substack{+3.1\\-2.6}$	\\[0.5ex]
	\hline
	\end{tabular}

	\label{MCFM_Differential_CrossSection}
\end{table*}

\section{Impact on the strange quark distribution in the proton}
\label{sec:QCD}

The associated \wc production at 13\TeV probes the strange quark distribution directly in the kinematic range of $\langle x \rangle \approx 0.007$ at the scale of $\mWsquared$.
The first measurement of a fiducial \wc cross section in $\Pp \Pp$ collisions was performed by the CMS experiment at a center-of-mass energy $\sqrts = 7\TeV$ with a total integrated luminosity of 5\fbinv~\cite{Chatrchyan:2013mza}.
The results were used in a QCD analysis~\cite{Chatrchyan:2013uja} together with measurements of neutral- and charged-current cross sections of DIS at HERA~\cite{Aaron:2009aa} and of the lepton charge asymmetry in $\PW$ production at $\sqrts = 7\TeV$ at the LHC~\cite{Chatrchyan:2013mza}.

The present measurement of the \wc production cross section at 13\TeV, determined as a function of the absolute pseudorapidity $\etaMuon$ of the muon from the $\PW$ boson decay and $\ptMuon > 26\GeV$, is used in an NLO QCD analysis.
This analysis also includes an updated combination of the inclusive DIS cross sections~\cite{Abramowicz:2015mha} and the available CMS measurements of the lepton charge asymmetry in $\PW$ boson production at $\sqrts = 7\TeV$~\cite{Chatrchyan:2013mza} and at $\sqrts = 8\TeV$~\cite{Khachatryan:2016pev}.
These latter measurements probe the valence quark distributions in the kinematic range $10^{-3} \leq x \leq 10^{-1}$ and have indirect sensitivity to the strange quark distribution.
The earlier CMS measurement~\cite{Chatrchyan:2013uja} of \wc production at $\sqrts = 7\TeV$ is also used to exploit the strange quark sensitive measurements at CMS in a joint QCD analysis.
The correlations of the experimental uncertainties within each individual data set are taken into account, whereas the CMS data sets are treated as uncorrelated to each other.
In particular, the measurements of \wc production at 7 and 13\TeV are treated as uncorrelated because of the different methods of charm tagging and the differences in reconstruction and event selection in the two data sets.

The theoretical predictions for the muon charge asymmetry and for \wc production are calculated at NLO using the {\MCFM} program~\cite{Campbell:1999ah,Campbell:2010ff}, which is interfaced to \textsc {applgrid} 1.4.56~\cite{Carli:2010rw}.

Version 2.0.0 of the open-source QCD fit framework for PDF determination \textsc{xFitter}~\cite{Alekhin:2014irh, herafitter} is used with the parton distributions evolved using the Dokshitzer--Gribov--Lipatov--Altarelli--Parisi equations~\cite{Gribov:1972ri,Altarelli:1977zs,Curci:1980uw,Furmanski:1980cm,Moch:2004pa,Vogt:2004mw} at NLO, as implemented in the \textsc{qcdnum} 17-00/06 program~\cite{Botje:2010ay}.

The Thorne--Roberts~\cite{Thorne:2006qt,Martin:2009ad} general mass variable flavor number scheme at NLO is used for the treatment of heavy quark contributions with heavy quark masses $m_{\cPqb} = 4.5\GeV$ and $m_{\cPqc} = 1.5\GeV$, which correspond to the values used in the signal MC simulation in the cross section measurements.
The renormalization and factorization scales are set to $Q$, which denotes the four-momentum transfer for the case of the DIS data and the mass of the $\PW$ boson for the case of the muon charge asymmetry and the \wc measurement.
The strong coupling constant is set to $\alpha_s (m_{\cPZ}) = 0.118$.
The $Q^2$ range of HERA data is restricted to $Q^2 \geq Q^2_{\min} = 3.5\GeV^2$ to ensure the applicability of pQCD over the kinematic range of the fit.
The procedure for the determination of the PDFs follows the approach used in the earlier CMS analyses~\cite{Chatrchyan:2013mza, Khachatryan:2016pev}.
In the following, a similar PDF parameterization is used as in the most recent CMS QCD analysis~\cite{Khachatryan:2016pev} of inclusive $\PW$ boson production.

The parameterized PDFs are the gluon distribution, $x\Pg$, the valence quark distributions, $x\cPqu_v$, $x\cPqd_v$, the $\cPqu$-type, $x\cPaqu$, and $x\cPaqd$-type anti-quark distributions, with $x\cPqs$ ($x\cPaqs$) denoting the strange (anti-)quark distribution.
The initial scale of the QCD evolution is chosen as $Q_0^2 = 1.9\GeV^2$.
At this scale, the parton distributions are parameterized as:
\begin{align}
x \cPqu_\mathrm{v}(x) &= A_{\cPqu_{\mathrm{v}}} ~  x^{B_{\cPqu_{\mathrm{v}}}} ~ (1-x)^{C_{\cPqu_{\mathrm{v}}}} ~(1+E_{\cPqu_{\mathrm{v}}} x^2),
\label{eq:uv}\\
x \cPqd_\mathrm{v}(x) &= A_{\cPqd_{\mathrm{v}}} ~ x^{B_{\cPqd_{\mathrm{v}}}} ~ (1-x)^{C_{\cPqd_{\mathrm{v}}}},
\label{eq:dv}\\
x \cPaqu(x) &= A_{\cPaqu } ~ x^{B_{\cPaqu}} ~ (1-x)^{C_{\cPaqu}}   ~(1+E_{\cPaqu} x^2),
\label{eq:Ubar}\\
 x \cPaqd(x) &=  A_{\cPaqd} ~ x^{B_{\cPaqd}} ~ (1-x)^{C_{\cPaqd}},
\label{eq:dbar} \\
 x \cPaqs(x) &=  A_{\cPaqs} ~ x^{B_{\cPaqs}} ~ (1-x)^{C_{\cPaqs}},
\label{eq:sq} \\
x \cPg(x) &= A_{\cPg} ~ x^{B_{\cPg}} ~ (1-x)^{C_{\cPg}}.
\label{eq:g}
\end{align}

The normalization parameters $A_{\cPqu_{\mathrm{v}}}$, $A_{\cPqd_\mathrm{v}}$, $A_\cPg$ are
determined by the QCD sum rules, the $B$ parameter is responsible for small-$x$ behavior of the PDFs, and the parameter $C$ describes the shape of the distribution as $x \to 1$.
The strangeness fraction $f_\cPqs = \cPaqs/( \cPaqd + \cPaqs)$ is a free parameter in the fit.

The strange quark distribution is determined by fitting the free parameters in Eqs.~(\ref{eq:uv})--(\ref{eq:g}).
The constraint $A_{\cPaqu} = A_{\cPaqd}$ ensures the same normalization for $\cPaqu$ and $\cPaqd$ densities at $x \to 0$.
It is assumed that $x\cPqs = x\cPaqs$.

In the earlier CMS analysis~\cite{Chatrchyan:2013mza}, the assumption $B_{\cPaqu} = B_{\cPaqd}$ was applied.
An alternative assumption $B_{\cPaqu} \ne B_{\cPaqd}$ led to a significant change in the result, which was included in the parameterization uncertainty.
In the present analysis, the $B$ parameters of the light sea quarks are independent from each other, $B_{\cPaqu} \neq B_{\cPaqd} \neq B_{\cPaqs}$, following the suggestion of Ref.~\cite{Alekhin:2017olj}.

For all measured data, the predicted and measured cross sections together with their corresponding uncertainties are used to build a global $\chiSquare$, minimized to determine the initial PDF parameters~\cite{Alekhin:2014irh, herafitter}.
The quality of the overall fit can be judged based on the global $\chiSquare$ divided by the number of degrees of freedom, $n_{\mathrm{dof}}$.
For each data set included in the fit, a partial $\chiSquare$ divided by the number of measurements (data points), $n_{\mathrm{dp}}$, is provided.
The correlated part of $\chiSquare$ quantifies the influence of the correlated systematic uncertainties in the fit.
The global and partial $\chiSquare$ values for each data set are listed in Table~\ref{chi2_paper_table_newparam}, illustrating a general agreement among all the data sets.

\begin{table*}[h!]
  \topcaption{The partial $\chiSquare$ per number of data points, $n_{\mathrm{dp}}$, and the global $\chiSquare$ per number of degree of freedom, $n_{\mathrm{dof}}$, resulting from the PDF fit.}
  \centering
    \begin{tabular}{lllll} \hline\\[-2.2ex]
  Data set &    &  $\chiSquare / n_{\mathrm{dp}}$  \\[0.5ex]
      \hline
  HERA I+II charged current & $\Pep \Pp$ & 43 / 39  \\
  HERA I+II charged current & $\Pem \Pp$ & 57 / 42   \\
  HERA I+II neutral current & $\Pem \Pp$ & 218 / 159   \\
  HERA I+II neutral current & $\Pep \Pp$, $E_{\Pp}$ = 820\GeV & 69 / 70  \\
  HERA I+II neutral current & $\Pep \Pp$, $E_{\Pp}$ = 920\GeV & 448 / 377  \\
  HERA I+II neutral current & $\Pep \Pp$, $E_{\Pp}$ = 460\GeV & 216 / 204  \\
  HERA I+II neutral current & $\Pep \Pp$, $E_{\Pp}$ = 575\GeV & 220 / 254  \\
  CMS $\PW$ muon charge asymmetry 7\TeV & & 13 / 11  \\
  CMS $\PW$ muon charge asymmetry 8\TeV & & 4.2 / 11  \\
  CMS \wc 7\TeV & & 2.2 / 5  \\
  CMS \wc 13\TeV & & 2.1 / 5&   \\
  Correlated $\chiSquare$ &  & 87  \\[\cmsTabSkip]
  Total $\chiSquare$ / dof &  & 1385 / 1160  \\[0.5ex]
  \hline\\
    \end{tabular}
  \label{chi2_paper_table_newparam}
\end{table*}

\begin{figure*}[ht!]
\centering
   \includegraphics[width=\cmsFigWidth]{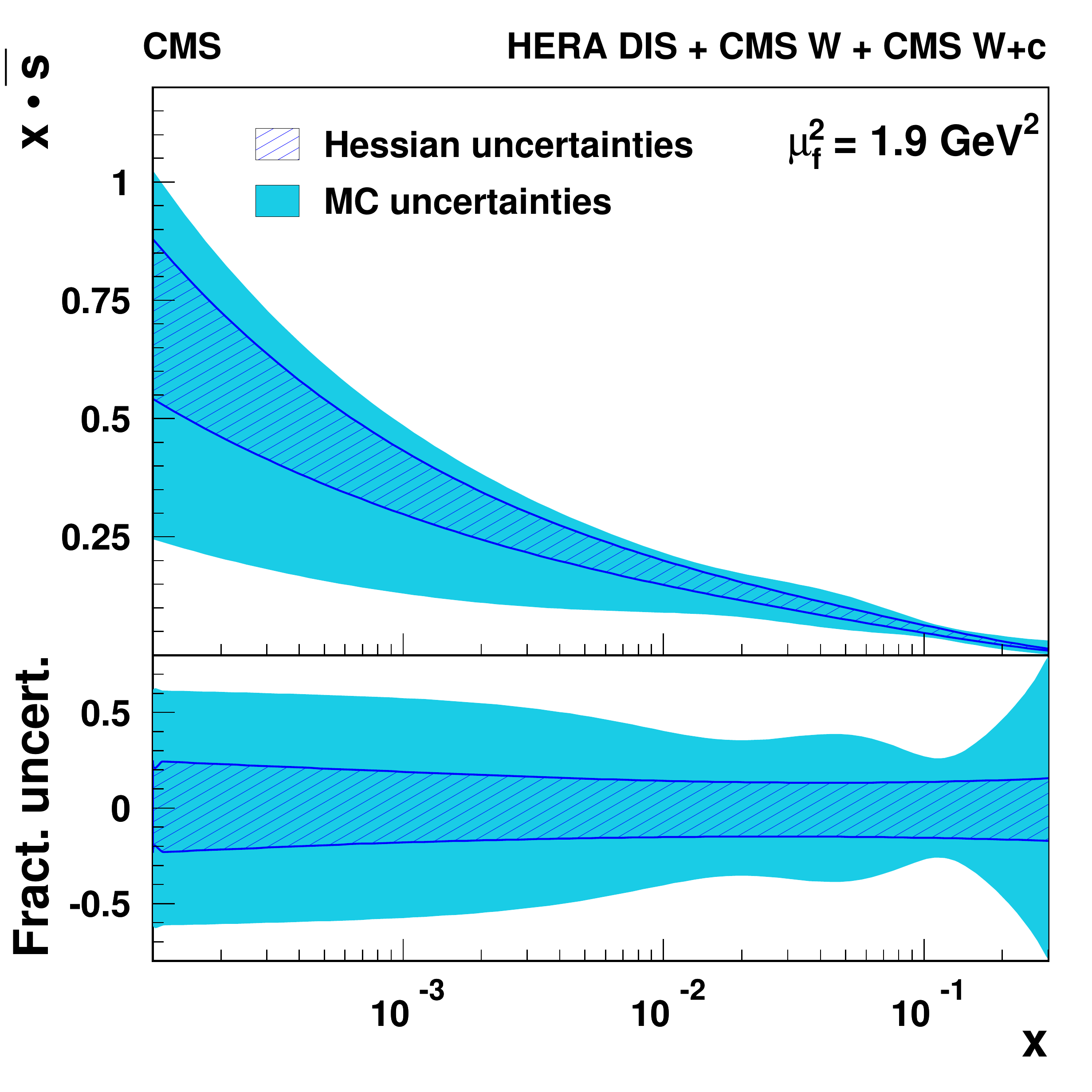}
   \includegraphics[width=\cmsFigWidth]{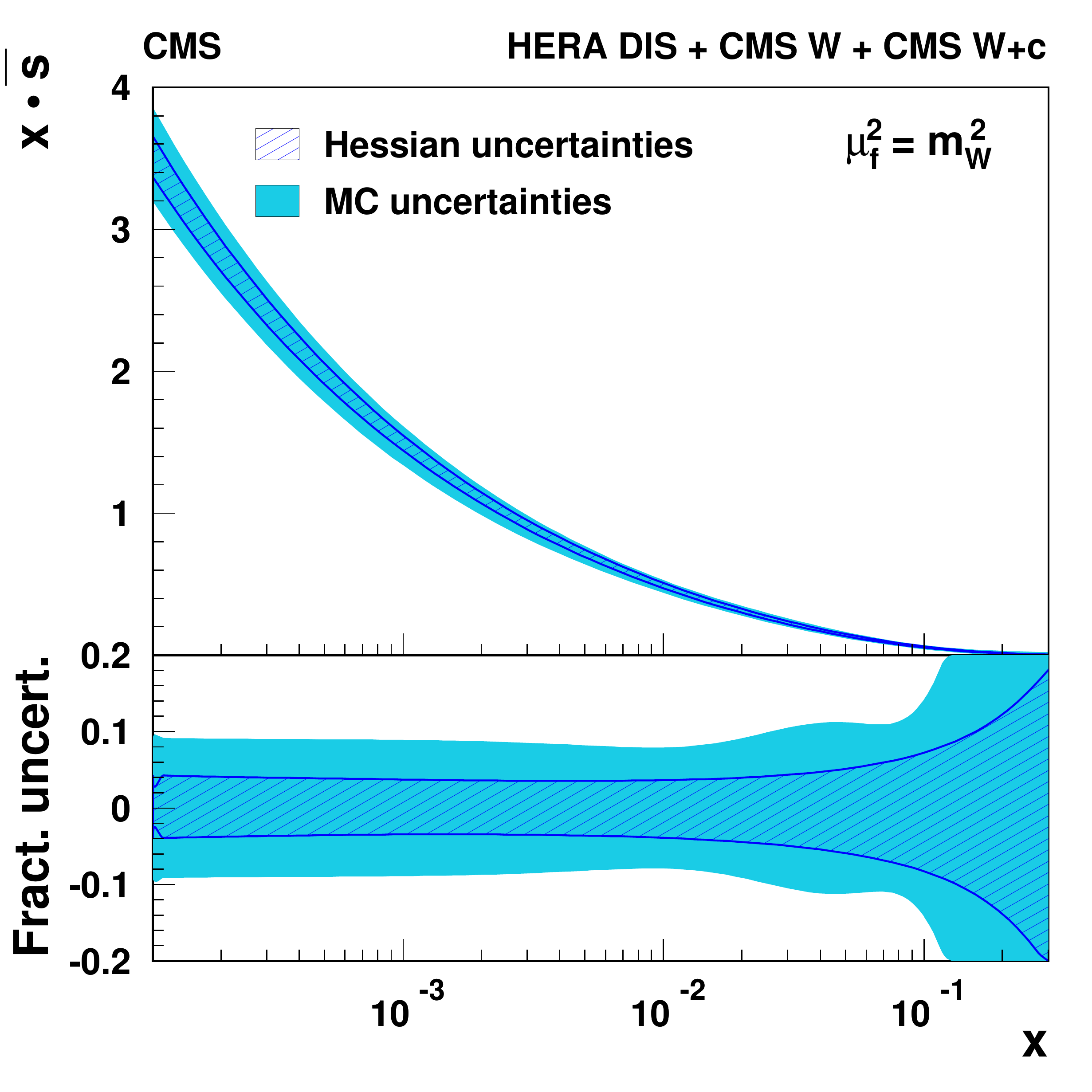}\\
  \includegraphics[width=\cmsFigWidth]{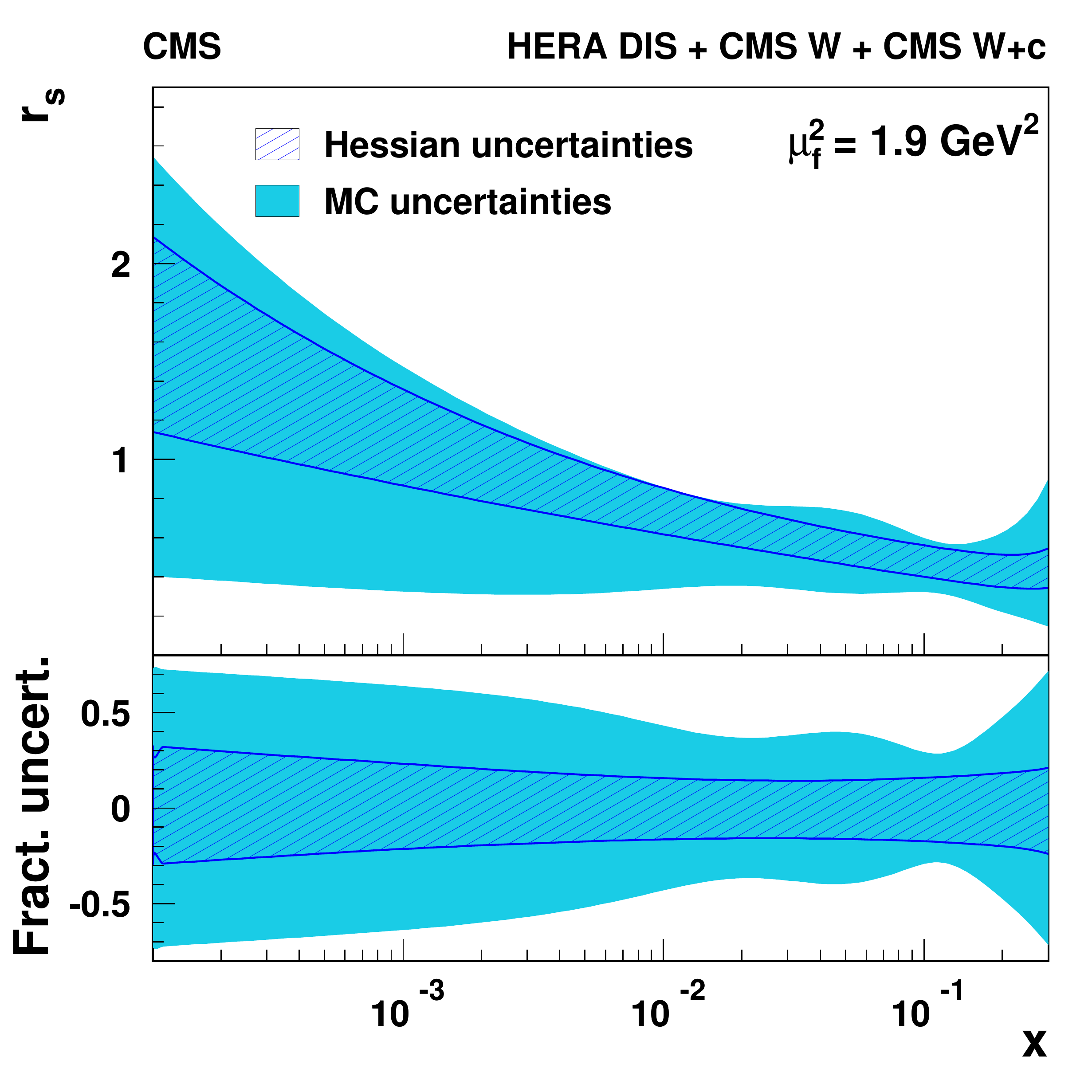}
   \includegraphics[width=\cmsFigWidth]{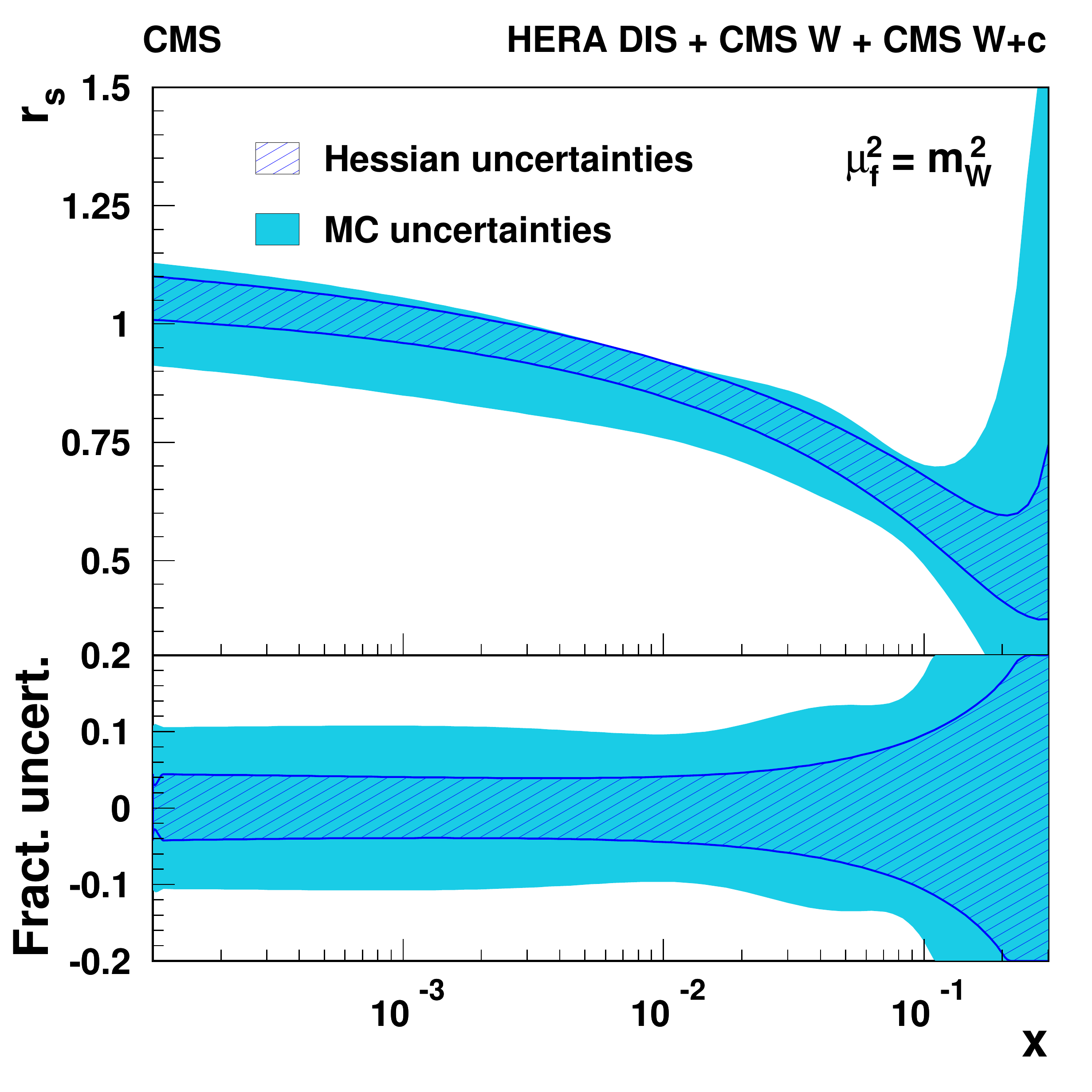}
\caption{The $\cPqs$ quark distribution (upper) and the strangeness suppression factor (lower)
as functions of $x$ at the factorization scale of 1.9\GeV$^2$ (left) and $\mWsquared$ (right).
The results of the current analysis are presented with the fit uncertainties estimated by the Hessian method (hatched band) and using MC replicas (shaded band).}
\label{s_rs_pdfs}
\end{figure*}

The PDF uncertainties are investigated according to the general approach of \textsc{HERAPDF} 1.0~\cite{Aaron:2009aa}.
The experimental PDF uncertainties arising from the uncertainties in the measurements are estimated by using the Hessian method~\cite{Pumplin:2001ct}, adopting the tolerance criterion of $\Delta\chiSquare = 1$.
The experimental uncertainties correspond to 68\% confidence level. Alternatively, the experimental uncertainties in the measurements are propagated to the extracted QCD fit parameters using the MC method~\cite{Giele:1998gw, Giele:2001mr}.
In this method, 426 replicas of pseudodata are generated, with measured values for the cross sections allowed to vary within the statistical and systematic uncertainties.
For each of them, the PDF fit is performed and the uncertainty is estimated as the root-mean-square around the central value.
Because of possible nonGaussian tails in the PDF uncertainties, the MC method is usually considered to be more robust and to give more realistic uncertainties, in particular for PDFs not strongly constrained by the measurements, \eg, in the case of too little or not very precise data.
In Fig.~\ref{s_rs_pdfs}, the distributions of the strange quark content $\cPqs(x,Q^2)$, and of the strangeness suppression factor $r_{\cPqs}(x,\PGm_f^2)=(\cPqs+\cPaqs)/(\cPaqu+\cPaqd)$ are presented.

\begin{figure*}[ht]
\centering
   \includegraphics[width=\cmsFigWidth]{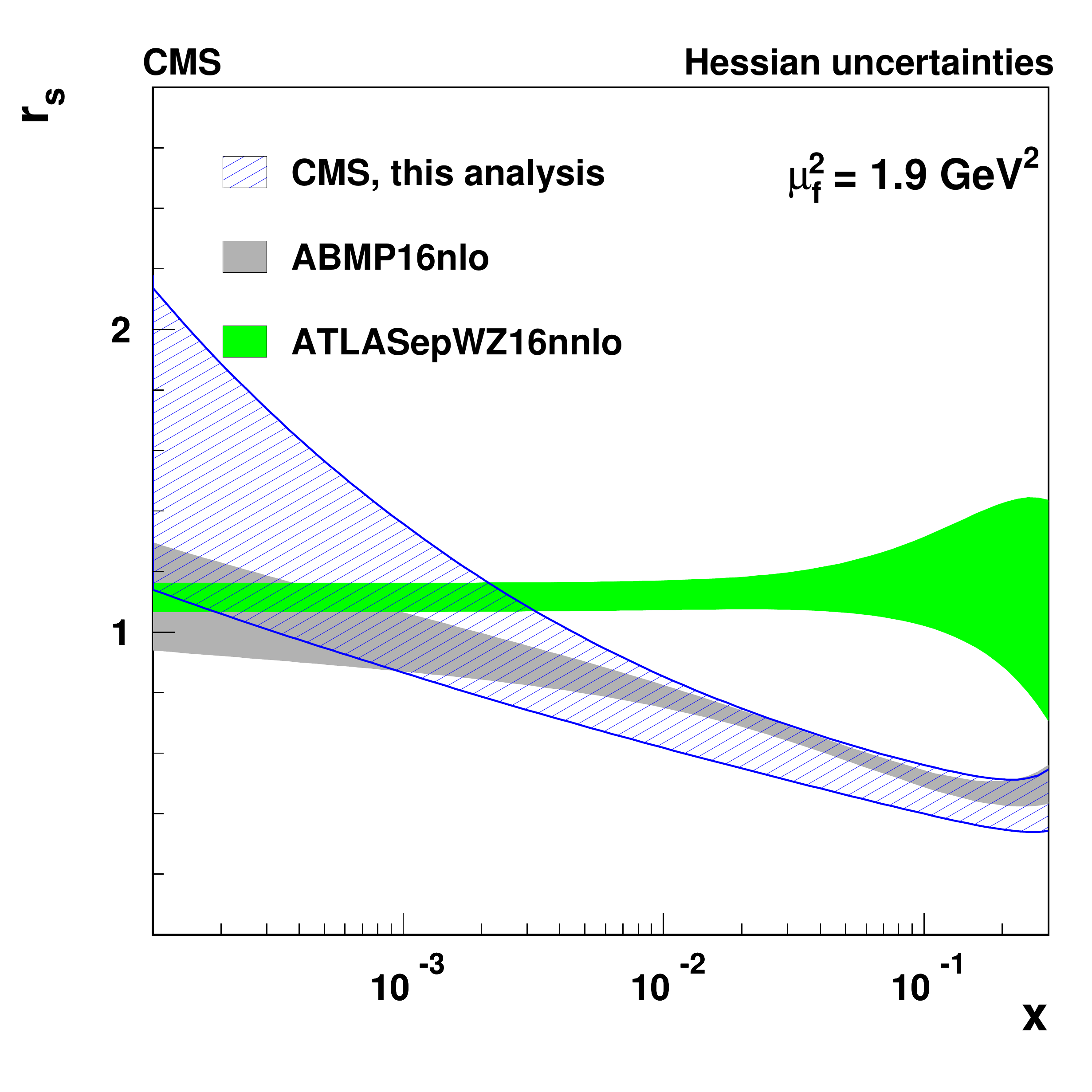}
  \includegraphics[width=\cmsFigWidth]{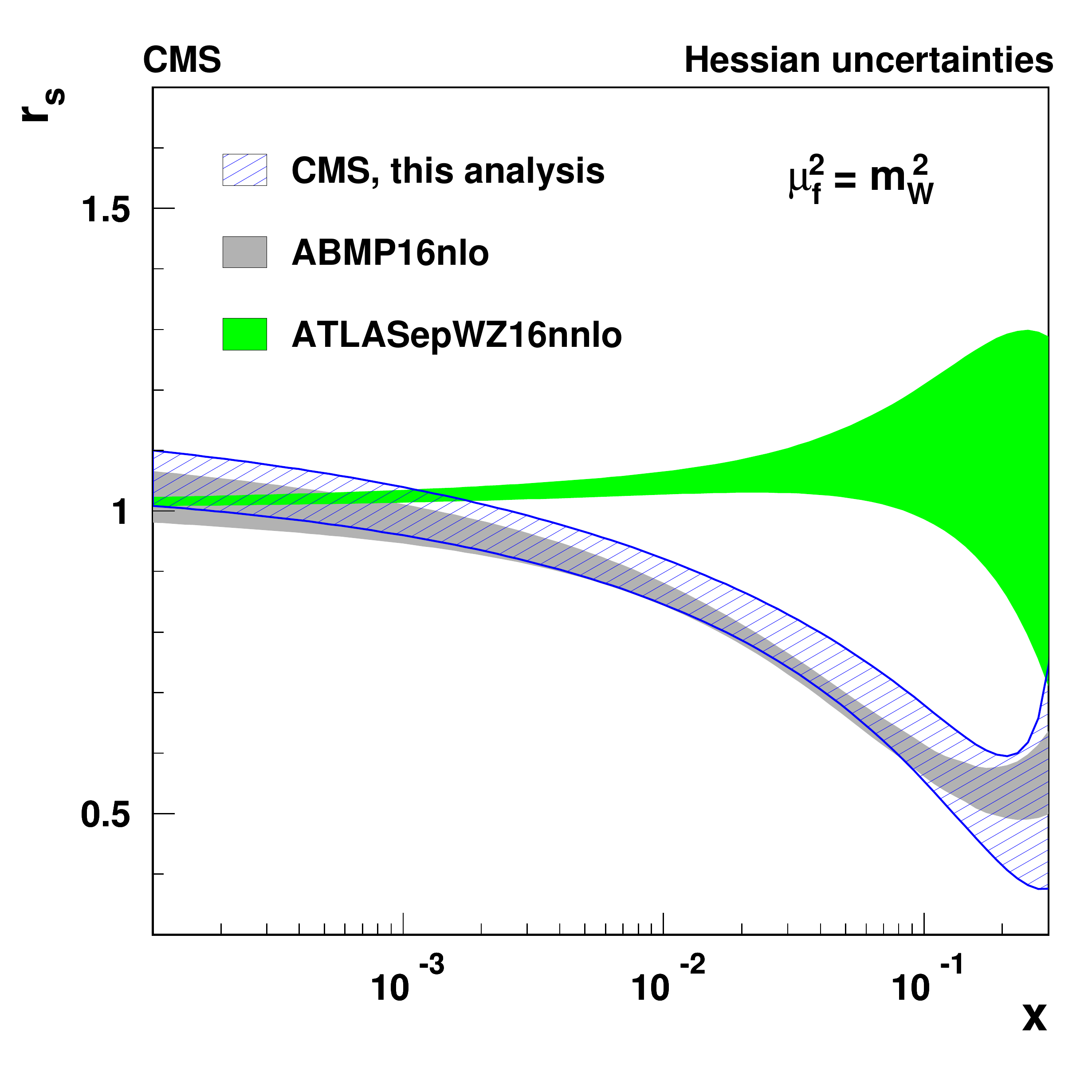}
\caption{The strangeness suppression factor as a function
of $x$ at the factorization scale of 1.9\GeV$^2$ (left) and $\mWsquared$ (right). The results of the current
analysis (hatched band) are compared to ABMP16nlo (dark shaded band) and ATLASepWZ16nnlo (light shaded band) PDFs.}
\label{rs_abmp_atlas}
\end{figure*}

\begin{figure*}[h!]
\centering
   \includegraphics[width=\cmsFigWidth]{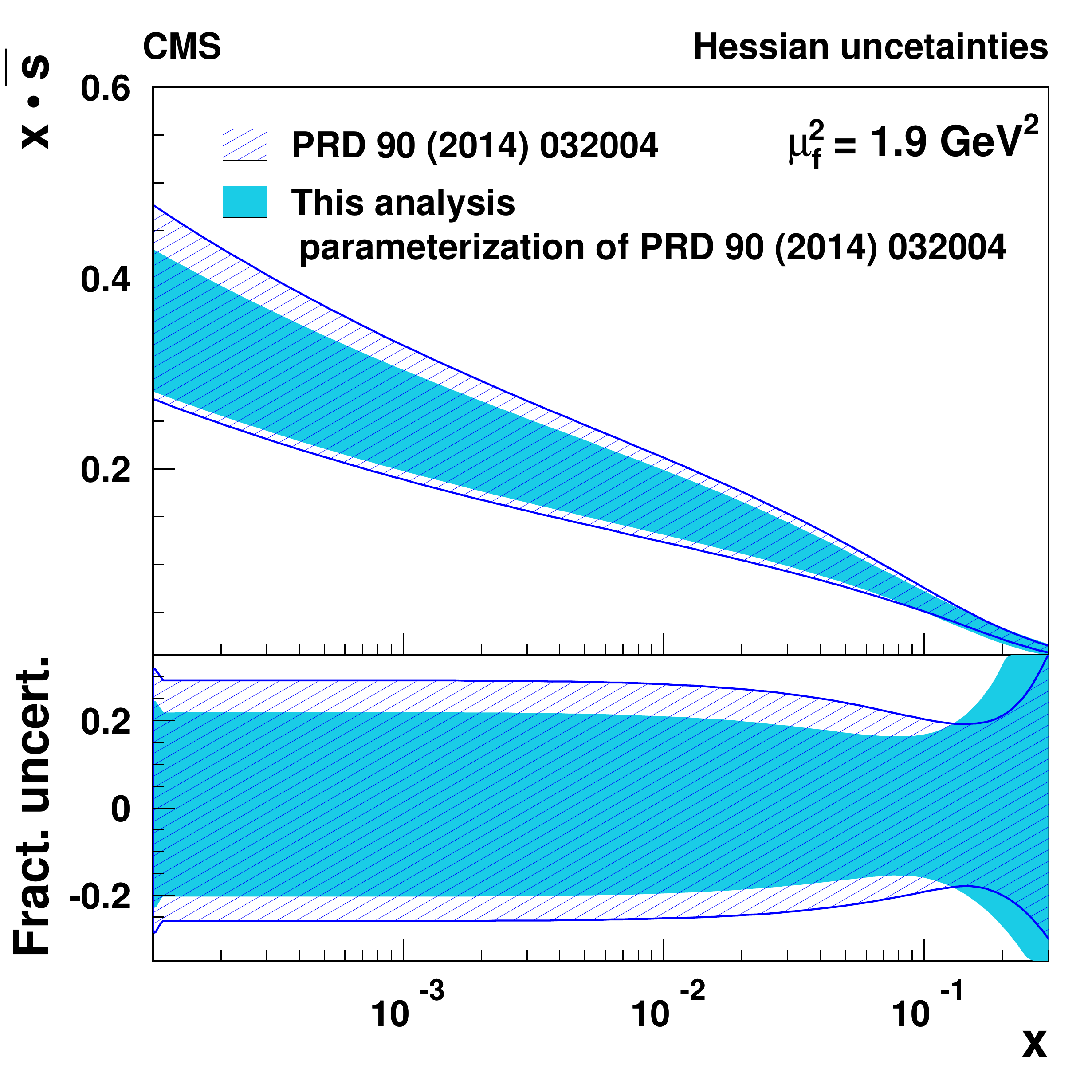}
   \includegraphics[width=\cmsFigWidth]{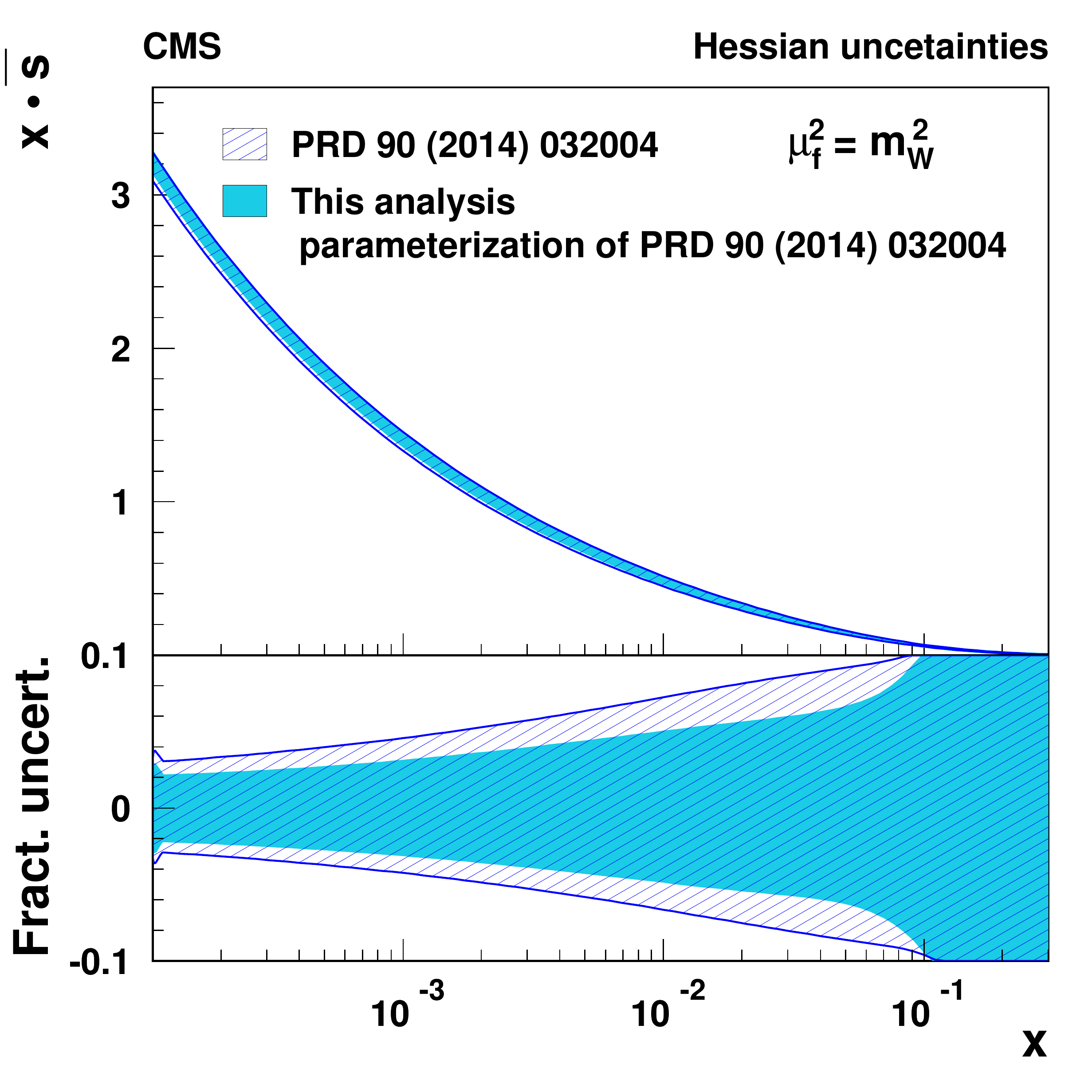}
\caption{The distributions of $\cPqs$ quarks (upper panel) in the proton and their relative uncertainty (lower panel) as a functions of $x$
at the factorization scale of 1.9\GeV$^2$ (left) and $\mWsquared$ (right). The result of the current analysis (filled band) is
compared to the result of Ref.~\cite{Chatrchyan:2013mza} (dashed band). The PDF uncertainties resulting from the fit are shown.}
\label{s-quark_pdf}
\end{figure*}

{\tolerance=900
In Fig.~\ref{rs_abmp_atlas} the strangeness suppression factor is shown in comparison with the ATLASepWZ16nnlo and the ABMP16nlo, similar to Fig.~1 in Ref.~\cite{Alekhin:2017olj}.
Whereas the CMS result for $\mathrm{r}_\mathrm{s}(x)$ is close to the ABMP16nlo PDF, it shows a significant difference with regard to the ATLASepWZ16nnlo PDF for $x > 10^{-3}$.
The significant excess of the strange quark content in the proton reported by ATLAS~\cite{Aaboud:2016btc} is not observed in the present analysis.
\par}

To investigate the impact of model assumptions on the resulting PDFs, alternative fits are performed,
in which the heavy quark masses are varied as $4.3\leq m_\cPqb\leq 5.0\GeV$, $1.37\leq m_\cPqc\leq 1.55\GeV$, and the value of $Q^2_{\min}$ imposed on the HERA data is varied in the interval $2.5 \leq Q^2_{\min}\leq 5.0\GeV^2$.
Also, the variations in PDF parameterization, following Ref.~\cite{Khachatryan:2016pev} are investigated.
These variations do not alter the results for the strange quark distribution or the suppression factor significantly, compared to the PDF fit uncertainty.
Since each global PDF group is using their own assumptions for the values of heavy quark masses and cutoffs on the DIS data, these model variations are not quantified further.

To compare the results of the present PDF fit with the earlier determination of the strange quark content in the proton at CMS~\cite{Chatrchyan:2013mza}, the ``free-s" parameterization of Ref.~\cite{Chatrchyan:2013mza} is used.
There, a flexible form~\cite{Thorne:2006qt,Martin:2009ad} for the gluon distribution was adopted, allowing the gluon to be negative.
Furthermore, the condition $B_{\cPaqu} = B_{\cPaqd} = B_{\cPaqs}$ was applied in the central parameterization, while $B_{\cPaqd} \neq B_{\cPaqs}$ was used to estimate the parameterization uncertainty.
A complete release of the condition  $B_{\cPaqu} = B_{\cPaqd} = B_{\cPaqs}$ was not possible because of limited data input, in contrast to the current analysis.
The same PDF parameterization was used in the ATLASepWZ16nnlo analysis~\cite{Aaboud:2016btc}.
The results are presented in Fig.~\ref{s-quark_pdf}.
The central value obtained of the $\cPqs$ quark distribution is well within the experimental uncertainty of the results  at $\sqrts = 7\TeV$, while the PDF uncertainty is reduced.

\section{Summary}
\label{Summary}
Associated production of $\PW$ bosons with charm quarks in proton-proton collisions at $\sqrts = 13\TeV$ is measured using the data collected by the CMS experiment in 2016 and corresponding to an integrated luminosity of 35.7\fbinv.
The $\PW$ boson is detected via the presence of a high-\pt muon and missing transverse momentum, suggesting the presence of a neutrino.
The charm quark is identified via the full reconstruction of the \Dstar meson decaying to $\PDz + \PiSlowpm \to \Kaonmp + \Pgppm + \PiSlowpm$.
Since in \wc production the $\PW$ boson and the $\cPqc$ quark have opposite charge, contributions from background processes, mainly $\cPqc$ quark production from gluon splitting, are largely removed by subtracting the events in which the charges of the $\PW$ boson and of the \Dstar meson have the same sign.
The fiducial cross sections are measured in the kinematic range of the muon transverse momentum $\ptMuon > 26\GeV$, pseudorapidity $\etaMuon < 2.4$, and transverse momentum of the charm quark $\ptCharm > 5\GeV$.
The  fiducial cross section of \wDstar production is measured in the kinematic range $\ptMuon > 26\GeV$, $\etaMuon < 2.4$, transverse momentum of the \Dstar meson $\ptDstar > 5\GeV$ and $\etaDstar < 2.4$, and compared to the Monte Carlo prediction.
The measurements are performed inclusively and in five bins of $\etaMuon$.

The obtained values for the inclusive fiducial \wc cross section and for the cross section ratio are:
\begin{equation}
	\sigmawc =  1026 \pm 31\stat\substack{+76\\-72}\syst\unit{pb},
	\label{Summary:AbsCrossSection}
\end{equation}
\begin{equation}
	\frac{\swpluscbar}{\swminusc} =  0.968 \pm 0.055\stat\substack{+0.015\\-0.028}\syst.
	\label{Summary:RatioCrossSection}
\end{equation}
The measurements are in good agreement with the theoretical predictions at next-to-leading order (NLO) for different sets of parton distribution functions (PDF), except for the one using the ATLASepWZ16nnlo PDF.
To illustrate the impact of these measurements in the determination of the strange quark distribution in the proton, the data is used in a QCD analysis at NLO together with inclusive DIS measurements and earlier results from CMS on \wc production and the lepton charge asymmetry in $\PW$ boson production.
The strange quark distribution and the strangeness suppression factor $r_{\cPqs}(x,\mu_f^2)=(\cPqs+\cPaqs)/(\cPaqu+\cPaqd)$ are determined and agree with earlier results obtained in neutrino scattering experiments.
The results do not support the hypothesis of an enhanced strange quark contribution in the proton quark sea reported by ATLAS~\cite{Aaboud:2016btc}.

\begin{acknowledgments}
We congratulate our colleagues in the CERN accelerator departments for the excellent performance of the LHC and thank the technical and administrative staffs at CERN and at other CMS institutes for their contributions to the success of the CMS effort. In addition, we gratefully acknowledge the computing centers and personnel of the Worldwide LHC Computing Grid for delivering so effectively the computing infrastructure essential to our analyses. Finally, we acknowledge the enduring support for the construction and operation of the LHC and the CMS detector provided by the following funding agencies: BMBWF and FWF (Austria); FNRS and FWO (Belgium); CNPq, CAPES, FAPERJ, FAPERGS, and FAPESP (Brazil); MES (Bulgaria); CERN; CAS, MoST, and NSFC (China); COLCIENCIAS (Colombia); MSES and CSF (Croatia); RPF (Cyprus); SENESCYT (Ecuador); MoER, ERC IUT, and ERDF (Estonia); Academy of Finland, MEC, and HIP (Finland); CEA and CNRS/IN2P3 (France); BMBF, DFG, and HGF (Germany); GSRT (Greece); NKFIA (Hungary); DAE and DST (India); IPM (Iran); SFI (Ireland); INFN (Italy); MSIP and NRF (Republic of Korea); MES (Latvia); LAS (Lithuania); MOE and UM (Malaysia); BUAP, CINVESTAV, CONACYT, LNS, SEP, and UASLP-FAI (Mexico); MOS (Montenegro); MBIE (New Zealand); PAEC (Pakistan); MSHE and NSC (Poland); FCT (Portugal); JINR (Dubna); MON, RosAtom, RAS, RFBR, and NRC KI (Russia); MESTD (Serbia); SEIDI, CPAN, PCTI, and FEDER (Spain); MOSTR (Sri Lanka); Swiss Funding Agencies (Switzerland); MST (Taipei); ThEPCenter, IPST, STAR, and NSTDA (Thailand); TUBITAK and TAEK (Turkey); NASU and SFFR (Ukraine); STFC (United Kingdom); DOE and NSF (USA).

 \hyphenation{Rachada-pisek} Individuals have received support from the Marie-Curie program and the European Research Council and Horizon 2020 Grant, contract No. 675440 (European Union); the Leventis Foundation; the A. P. Sloan Foundation; the Alexander von Humboldt Foundation; the Belgian Federal Science Policy Office; the Fonds pour la Formation \`a la Recherche dans l'Industrie et dans l'Agriculture (FRIA-Belgium); the Agentschap voor Innovatie door Wetenschap en Technologie (IWT-Belgium); the F.R.S.-FNRS and FWO (Belgium) under the ``Excellence of Science - EOS" - be.h project n. 30820817; the Ministry of Education, Youth and Sports (MEYS) of the Czech Republic; the Lend\"ulet (``Momentum") Programme and the J\'anos Bolyai Research Scholarship of the Hungarian Academy of Sciences, the New National Excellence Program \'UNKP, the NKFIA research grants 123842, 123959, 124845, 124850 and 125105 (Hungary); the Council of Science and Industrial Research, India; the HOMING PLUS program of the Foundation for Polish Science, cofinanced from European Union, Regional Development Fund, the Mobility Plus program of the Ministry of Science and Higher Education, the National Science Center (Poland), contracts Harmonia 2014/14/M/ST2/00428, Opus 2014/13/B/ST2/02543, 2014/15/B/ST2/03998, and 2015/19/B/ST2/02861, Sonata-bis 2012/07/E/ST2/01406; the National Priorities Research Program by Qatar National Research Fund; the Programa Estatal de Fomento de la Investigaci{\'o}n Cient{\'i}fica y T{\'e}cnica de Excelencia Mar\'{\i}a de Maeztu, grant MDM-2015-0509 and the Programa Severo Ochoa del Principado de Asturias; the Thalis and Aristeia programs cofinanced by EU-ESF and the Greek NSRF; the Rachadapisek Sompot Fund for Postdoctoral Fellowship, Chulalongkorn University and the Chulalongkorn Academic into Its 2nd Century Project Advancement Project (Thailand); the Welch Foundation, contract C-1845; and the Weston Havens Foundation (USA).
\end{acknowledgments}

\bibliography{auto_generated}

\cleardoublepage \appendix\section{The CMS Collaboration \label{app:collab}}\begin{sloppypar}\hyphenpenalty=5000\widowpenalty=500\clubpenalty=5000\input{SMP-17-014-authorlist.tex}\end{sloppypar}
\end{document}

%% file: SMP-17-014-authorlist.tex
\vskip\cmsinstskip
\textbf{Yerevan Physics Institute, Yerevan, Armenia}\\*[0pt]
A.M.~Sirunyan, A.~Tumasyan
\vskip\cmsinstskip
\textbf{Institut f\"{u}r Hochenergiephysik, Wien, Austria}\\*[0pt]
W.~Adam, F.~Ambrogi, E.~Asilar, T.~Bergauer, J.~Brandstetter, M.~Dragicevic, J.~Er\"{o}, A.~Escalante~Del~Valle, M.~Flechl, R.~Fr\"{u}hwirth\cmsAuthorMark{1}, V.M.~Ghete, J.~Hrubec, M.~Jeitler\cmsAuthorMark{1}, N.~Krammer, I.~Kr\"{a}tschmer, D.~Liko, T.~Madlener, I.~Mikulec, N.~Rad, H.~Rohringer, J.~Schieck\cmsAuthorMark{1}, R.~Sch\"{o}fbeck, M.~Spanring, D.~Spitzbart, A.~Taurok, W.~Waltenberger, J.~Wittmann, C.-E.~Wulz\cmsAuthorMark{1}, M.~Zarucki
% ...
\vskip\cmsinstskip
% ...
1:  Also at Vienna University of Technology, Vienna, Austria\\
% ...


%% file: SMP-17-014_temp.bbl
\providecommand{\href}[2]{#2}\begingroup\raggedright\begin{thebibliography}{10}%
\makeatletter
\providecommand{\hrefCMSnoop }[0]{\@secondoftwo}%
\makeatother
\providecommand{\doi}{\texttt{doi:}\begingroup \urlstyle{tt}\Url}

\bibitem{Aaboud:2017svj}
\hrefCMSnoop {}{{ATLAS Collaboration}, ``{Measurement of the $W$-boson mass in
  pp collisions at \newline $\sqrt{s}=7$ TeV with the ATLAS detector}'',}
  \textit{ Eur. Phys. J. C} \textbf{ 78} (2018) 110,
  \href{http://dx.doi.org/10.1140/epjc/s10052-017-5475-4}{\doi{10.1140/epjc/s10052-017-5475-4}},
\href{http://www.arXiv.org/abs/1701.07240}{\texttt{arXiv:1701.07240}}.

\bibitem{Goncharov:2001qe}
\hrefCMSnoop {}{{NuTeV} Collaboration, ``{Precise measurement of dimuon
  production cross-sections in muon neutrino Fe and muon anti-neutrino Fe deep
  inelastic scattering at the Tevatron}'',} \textit{ Phys. Rev. D} \textbf{ 64}
  (2001) 112006,
  \href{http://dx.doi.org/10.1103/PhysRevD.64.112006}{\doi{10.1103/PhysRevD.64.112006}},
\href{http://www.arXiv.org/abs/hep-ex/0102049}{\texttt{arXiv:hep-ex/0102049}}.

\bibitem{Bazarko:1994tt}
\hrefCMSnoop {}{{CCFR} Collaboration, ``{Determination of the strange quark
  content of the nucleon from a next-to-leading order QCD analysis of neutrino
  charm production}'',} \textit{ Z. Phys. C} \textbf{ 65} (1995) 189,
  \href{http://dx.doi.org/10.1007/BF01571875}{\doi{10.1007/BF01571875}},
\href{http://www.arXiv.org/abs/hep-ex/9406007}{\texttt{arXiv:hep-ex/9406007}}.

\bibitem{Samoylov:2013xoa}
\hrefCMSnoop {}{{NOMAD} Collaboration, ``{A precision measurement of charm
  dimuon production in neutrino interactions from the NOMAD experiment}'',}
  \textit{ Nucl. Phys. B} \textbf{ 876} (2013) 339,
  \href{http://dx.doi.org/10.1016/j.nuclphysb.2013.08.021}{\doi{10.1016/j.nuclphysb.2013.08.021}},
\href{http://www.arXiv.org/abs/1308.4750}{\texttt{arXiv:1308.4750}}.

\bibitem{KayisTopaksu:2011mx}
\hrefCMSnoop {}{A.~Kayis-Topaksu {et~al.}, ``{Measurement of charm production
  in neutrino charged-current interactions}'',} \textit{ New J. Phys.} \textbf{
  13} (2011) 093002,
  \href{http://dx.doi.org/10.1088/1367-2630/13/9/093002}{\doi{10.1088/1367-2630/13/9/093002}},
\href{http://www.arXiv.org/abs/1107.0613}{\texttt{arXiv:1107.0613}}.

\bibitem{Aad:2012sb}
\hrefCMSnoop {}{{ATLAS Collaboration}, ``{Determination of the strange quark
  density of the proton from ATLAS measurements of the $W \to \ell \nu$ and $Z
  \to \ell\ell$ cross sections}'',} \textit{ Phys. Rev. Lett.} \textbf{ 109}
  (2012) 012001,
  \href{http://dx.doi.org/10.1103/PhysRevLett.109.012001}{\doi{10.1103/PhysRevLett.109.012001}},
\href{http://www.arXiv.org/abs/1203.4051}{\texttt{arXiv:1203.4051}}.

\bibitem{Aaltonen:2007dm}
\hrefCMSnoop {}{{CDF} Collaboration, ``{First measurement of the production of
  a $\PW$ boson in association with a single charm quark in $\mathrm{p}
  \bar{\mathrm{p}}$ collisions at $\sqrt{s}$ = 1.96 TeV}'',} \textit{ Phys.
  Rev. Lett.} \textbf{ 100} (2008) 091803,
  \href{http://dx.doi.org/10.1103/PhysRevLett.100.091803}{\doi{10.1103/PhysRevLett.100.091803}},
\href{http://www.arXiv.org/abs/0711.2901}{\texttt{arXiv:0711.2901}}.

\bibitem{Aaltonen:2012wn}
\hrefCMSnoop {}{{CDF} Collaboration, ``{Observation of the production of a W
  boson in association with a single charm quark}'',} \textit{ Phys. Rev.
  Lett.} \textbf{ 110} (2013) 071801,
  \href{http://dx.doi.org/10.1103/PhysRevLett.110.071801}{\doi{10.1103/PhysRevLett.110.071801}},
\href{http://www.arXiv.org/abs/1209.1921}{\texttt{arXiv:1209.1921}}.

\bibitem{Abazov:2008qz}
\hrefCMSnoop {}{{D0} Collaboration, ``{Measurement of the ratio of the $\Pp
  \Pap \to \PW + \cPqc$- jet cross section to the inclusive $\Pp \Pap \to \PW
  +$ jets cross section}'',} \textit{ Phys. Lett. B} \textbf{ 666} (2008) 23,
  \href{http://dx.doi.org/10.1016/j.physletb.2008.06.067}{\doi{10.1016/j.physletb.2008.06.067}},
\href{http://www.arXiv.org/abs/0803.2259}{\texttt{arXiv:0803.2259}}.

\bibitem{Chatrchyan:2013uja}
\hrefCMSnoop {}{{CMS Collaboration}, ``{Measurement of associated W+charm
  production in pp collisions at $\sqrt{s}$ = 7 TeV}'',} \textit{ JHEP}
  \textbf{ 02} (2014) 013,
  \href{http://dx.doi.org/10.1007/JHEP02(2014)013}{\doi{10.1007/JHEP02(2014)013}},
\href{http://www.arXiv.org/abs/1310.1138}{\texttt{arXiv:1310.1138}}.

\bibitem{Chatrchyan:2013mza}
\hrefCMSnoop {}{{CMS Collaboration}, ``{Measurement of the muon charge
  asymmetry in inclusive $\mathrm{pp} \to \mathrm{W}+X$ production at $\sqrt s
  =$ 7 TeV and an improved determination of light parton distribution
  functions}'',} \textit{ Phys. Rev. D} \textbf{ 90} (2014) 032004,
  \href{http://dx.doi.org/10.1103/PhysRevD.90.032004}{\doi{10.1103/PhysRevD.90.032004}},
\href{http://www.arXiv.org/abs/1312.6283}{\texttt{arXiv:1312.6283}}.

\bibitem{Aad:2014xca}
\hrefCMSnoop {}{{ATLAS Collaboration}, ``{Measurement of the production of a
  $\PW$ boson in association with a charm quark in $\Pp\Pp$ collisions at
  $\sqrt{s} =$ 7 TeV with the ATLAS detector}'',} \textit{ JHEP} \textbf{ 05}
  (2014) 068,
  \href{http://dx.doi.org/10.1007/JHEP05(2014)068}{\doi{10.1007/JHEP05(2014)068}},
\href{http://www.arXiv.org/abs/1402.6263}{\texttt{arXiv:1402.6263}}.

\bibitem{Alekhin:2014sya}
S.~Alekhin\hrefCMSnoop {}{ {et~al.}, ``{Determination of strange sea quark
  distributions from fixed-target and collider data}'',} \textit{ Phys. Rev. D}
  \textbf{ 91} (2015) 094002,
  \href{http://dx.doi.org/10.1103/PhysRevD.91.094002}{\doi{10.1103/PhysRevD.91.094002}},
\href{http://www.arXiv.org/abs/1404.6469}{\texttt{arXiv:1404.6469}}.

\bibitem{Aaboud:2016btc}
\hrefCMSnoop {}{{ATLAS Collaboration}, ``{Precision measurement and
  interpretation of inclusive $W^+$ , $W^-$ and $Z/\gamma ^*$ production cross
  sections with the ATLAS detector}'',} \textit{ Eur. Phys. J. C} \textbf{ 77}
  (2017) 367,
  \href{http://dx.doi.org/10.1140/epjc/s10052-017-4911-9}{\doi{10.1140/epjc/s10052-017-4911-9}},
\href{http://www.arXiv.org/abs/1612.03016}{\texttt{arXiv:1612.03016}}.

\bibitem{Alekhin:2017olj}
\hrefCMSnoop {}{S.~Alekhin, J.~Bl{\"u}mlein, and S.~Moch, ``{Strange sea
  determination from collider data}'',} \textit{ Phys. Lett. B} \textbf{ 777}
  (2018) 134,
  \href{http://dx.doi.org/10.1016/j.physletb.2017.12.024}{\doi{10.1016/j.physletb.2017.12.024}},
\href{http://www.arXiv.org/abs/1708.01067}{\texttt{arXiv:1708.01067}}.

\bibitem{Aaij:2015cha}
\hrefCMSnoop {}{{LHCb Collaboration}, ``{Study of $\PW$ boson production in
  association with beauty and charm}'',} \textit{ Phys. Rev. D} \textbf{ 92}
  (2015) 052001,
  \href{http://dx.doi.org/10.1103/PhysRevD.92.052001}{\doi{10.1103/PhysRevD.92.052001}},
\href{http://www.arXiv.org/abs/1505.04051}{\texttt{arXiv:1505.04051}}.

\bibitem{Campbell:1999ah}
\hrefCMSnoop {}{J.~M. Campbell and R.~K. Ellis, ``{Update on vector boson pair
  production at hadron colliders}'',} \textit{ Phys. Rev. D} \textbf{ 60}
  (1999) 113006,
  \href{http://dx.doi.org/10.1103/PhysRevD.60.113006}{\doi{10.1103/PhysRevD.60.113006}},
\href{http://www.arXiv.org/abs/hep-ph/9905386}{\texttt{arXiv:hep-ph/9905386}}.

\bibitem{Campbell:2010ff}
\hrefCMSnoop {}{J.~M. Campbell and R.~K. Ellis, ``{MCFM for the Tevatron and
  the LHC}'',} \textit{ Nucl. Phys. Proc. Suppl.} \textbf{ 205-206} (2010) 10,
  \href{http://dx.doi.org/10.1016/j.nuclphysbps.2010.08.011}{\doi{10.1016/j.nuclphysbps.2010.08.011}},
\href{http://www.arXiv.org/abs/1007.3492}{\texttt{arXiv:1007.3492}}.

\bibitem{Campbell:2012uf}
\hrefCMSnoop {}{J.~M. Campbell and R.~K. Ellis, ``{Top quark processes at NLO
  in production and decay}'',} \textit{ J. Phys. G} \textbf{ 42} (2015) 015005,
  \href{http://dx.doi.org/10.1088/0954-3899/42/1/015005}{\doi{10.1088/0954-3899/42/1/015005}},
\href{http://www.arXiv.org/abs/1204.1513}{\texttt{arXiv:1204.1513}}.

\bibitem{Chatrchyan:2014fea}
\hrefCMSnoop {}{{CMS Collaboration}, ``{Description and performance of track
  and primary vertex reconstruction with the CMS tracker}'',} \textit{ JINST}
  \textbf{ 9} (2014) P10009,
  \href{http://dx.doi.org/10.1088/1748-0221/9/10/P10009}{\doi{10.1088/1748-0221/9/10/P10009}},
\href{http://www.arXiv.org/abs/1405.6569}{\texttt{arXiv:1405.6569}}.

\bibitem{Cacciari:2008gp}
\hrefCMSnoop {}{M.~Cacciari, G.~P. Salam, and G.~Soyez, ``{The anti-$\kt$ jet
  clustering algorithm}'',} \textit{ JHEP} \textbf{ 04} (2008) 063,
  \href{http://dx.doi.org/10.1088/1126-6708/2008/04/063}{\doi{10.1088/1126-6708/2008/04/063}},
  \href{http://www.arXiv.org/abs/0802.1189}{\texttt{arXiv:0802.1189}}.

\bibitem{Cacciari:2011ma}
\hrefCMSnoop {}{M.~Cacciari, G.~P. Salam, and G.~Soyez, ``{FastJet user
  manual}'',} \textit{ Eur. Phys. J. C} \textbf{ 72} (2012) 1896,
  \href{http://dx.doi.org/10.1140/epjc/s10052-012-1896-2}{\doi{10.1140/epjc/s10052-012-1896-2}},
\href{http://www.arXiv.org/abs/1111.6097}{\texttt{arXiv:1111.6097}}.

\bibitem{Sirunyan:2018fpa}
\hrefCMSnoop {}{{CMS Collaboration}, ``{Performance of the CMS muon detector
  and muon reconstruction with proton-proton collisions at $\sqrt{s}=$ 13
  TeV}'',} \textit{ JINST} \textbf{ 13} (2018) P06015,
  \href{http://dx.doi.org/10.1088/1748-0221/13/06/P06015}{\doi{10.1088/1748-0221/13/06/P06015}},
\href{http://www.arXiv.org/abs/1804.04528}{\texttt{arXiv:1804.04528}}.

\bibitem{Chatrchyan:2008zzk}
\hrefCMSnoop {}{{CMS Collaboration}, ``The {CMS} experiment at the {CERN}
  {LHC}'',} \textit{ JINST} \textbf{ 3} (2008) S08004,
\href{http://dx.doi.org/10.1088/1748-0221/3/08/S08004}{\doi{10.1088/1748-0221/3/08/S08004}}.

\bibitem{Khachatryan:2016bia}
\hrefCMSnoop {}{{CMS Collaboration}, ``{The CMS trigger system}'',} \textit{
  JINST} \textbf{ 12} (2017) P01020,
  \href{http://dx.doi.org/10.1088/1748-0221/12/01/P01020}{\doi{10.1088/1748-0221/12/01/P01020}},
\href{http://www.arXiv.org/abs/1609.02366}{\texttt{arXiv:1609.02366}}.

\bibitem{Sirunyan:2017ulk}
\hrefCMSnoop {}{{CMS Collaboration}, ``{Particle-flow reconstruction and global
  event description with the CMS detector}'',} \textit{ JINST} \textbf{ 12}
  (2017) P10003,
  \href{http://dx.doi.org/10.1088/1748-0221/12/10/P10003}{\doi{10.1088/1748-0221/12/10/P10003}},
\href{http://www.arXiv.org/abs/1706.04965}{\texttt{arXiv:1706.04965}}.

\bibitem{Agostinelli:2002hh}
\hrefCMSnoop {}{{GEANT4} Collaboration, ``{\GEANTfour}---a simulation
  toolkit'',} \textit{ Nucl. Instrum. Meth. A} \textbf{ 506} (2003) 250,
\href{http://dx.doi.org/10.1016/S0168-9002(03)01368-8}{\doi{10.1016/S0168-9002(03)01368-8}}.

\bibitem{Alwall:2014hca}
J.~Alwall\hrefCMSnoop {}{ {et~al.}, ``{The automated computation of tree-level
  and next-to-leading order differential cross sections, and their matching to
  parton shower simulations}'',} \textit{ JHEP} \textbf{ 07} (2014) 079,
  \href{http://dx.doi.org/10.1007/JHEP07(2014)079}{\doi{10.1007/JHEP07(2014)079}},
\href{http://www.arXiv.org/abs/1405.0301}{\texttt{arXiv:1405.0301}}.

\bibitem{Sjostrand:2007gs}
\hrefCMSnoop {}{T.~Sj{\"o}strand, S.~Mrenna, and P.~Z. Skands, ``{A brief
  introduction to PYTHIA 8.1}'',} \textit{ Comput. Phys. Commun.} \textbf{ 178}
  (2008) 852,
  \href{http://dx.doi.org/10.1016/j.cpc.2008.01.036}{\doi{10.1016/j.cpc.2008.01.036}},
\href{http://www.arXiv.org/abs/0710.3820}{\texttt{arXiv:0710.3820}}.

\bibitem{Frederix:2012ps}
\hrefCMSnoop {}{R.~Frederix and S.~Frixione, ``{Merging meets matching in
  MC@NLO}'',} \textit{ JHEP} \textbf{ 12} (2012) 061,
  \href{http://dx.doi.org/10.1007/JHEP12(2012)061}{\doi{10.1007/JHEP12(2012)061}},
\href{http://www.arXiv.org/abs/1209.6215}{\texttt{arXiv:1209.6215}}.

\bibitem{Ball:2014uwa}
\hrefCMSnoop {}{{NNPDF} Collaboration, ``{Parton distributions for the LHC run
  II}'',} \textit{ JHEP} \textbf{ 04} (2015) 040,
  \href{http://dx.doi.org/10.1007/JHEP04(2015)040}{\doi{10.1007/JHEP04(2015)040}},
\href{http://www.arXiv.org/abs/1410.8849}{\texttt{arXiv:1410.8849}}.

\bibitem{Campbell:2014kua}
\hrefCMSnoop {}{J.~M. Campbell, R.~K. Ellis, P.~Nason, and E.~Re, ``{Top-pair
  production and decay at NLO matched with parton showers}'',} \textit{ JHEP}
  \textbf{ 04} (2015) 114,
  \href{http://dx.doi.org/10.1007/JHEP04(2015)114}{\doi{10.1007/JHEP04(2015)114}},
\href{http://www.arXiv.org/abs/1412.1828}{\texttt{arXiv:1412.1828}}.

\bibitem{Frederix:2012dh}
\hrefCMSnoop {}{R.~Frederix, E.~Re, and P.~Torrielli, ``{Single-top $t$-channel
  hadroproduction in the four-flavour scheme with POWHEG and aMC@NLO}'',}
  \textit{ JHEP} \textbf{ 09} (2012) 130,
  \href{http://dx.doi.org/10.1007/JHEP09(2012)130}{\doi{10.1007/JHEP09(2012)130}},
\href{http://www.arXiv.org/abs/1207.5391}{\texttt{arXiv:1207.5391}}.

\bibitem{Alioli:2009je}
\hrefCMSnoop {}{S.~Alioli, P.~Nason, C.~Oleari, and E.~Re, ``{NLO single-top
  production matched with shower in POWHEG: $s$- and $t$-channel
  contributions}'',} \textit{ JHEP} \textbf{ 09} (2009) 111,
  \href{http://dx.doi.org/10.1088/1126-6708/2009/09/111}{\doi{10.1088/1126-6708/2009/09/111}},
  \href{http://www.arXiv.org/abs/0907.4076}{\texttt{arXiv:0907.4076}}.
[Erratum: {\it JHEP\/} {\bf 02} (2010) 011].

\bibitem{Re:2010bp}
\hrefCMSnoop {}{E.~Re, ``{Single-top \PW\cPqt-channel production matched with
  parton showers using the POWHEG method}'',} \textit{ Eur. Phys. J. C}
  \textbf{ 71} (2011) 1547,
  \href{http://dx.doi.org/10.1140/epjc/s10052-011-1547-z}{\doi{10.1140/epjc/s10052-011-1547-z}},
\href{http://www.arXiv.org/abs/1009.2450}{\texttt{arXiv:1009.2450}}.

\bibitem{Khachatryan:2015pea}
\hrefCMSnoop {}{{CMS Collaboration}, ``{Event generator tunes obtained from
  underlying event and multiparton scattering measurements}'',} \textit{ Eur.
  Phys. J. C} \textbf{ 76} (2016) 155,
  \href{http://dx.doi.org/10.1140/epjc/s10052-016-3988-x}{\doi{10.1140/epjc/s10052-016-3988-x}},
\href{http://www.arXiv.org/abs/1512.00815}{\texttt{arXiv:1512.00815}}.

\bibitem{CMS-PAS-TOP-16-021}
\hrefCMSnoop {}{{CMS Collaboration}, ``{Investigations of the impact of the
  parton shower tuning in Pythia 8 in the modelling of $\mathrm{t\overline{t}}$
  at $\sqrt{s}=8$ and 13 TeV}'',} CMS Physics Analysis Summary
  CMS-PAS-TOP-16-021, 2016.
\newblock \url{https://cds.cern.ch/record/2235192}.

\bibitem{Buckley:2010ar}
A.~Buckley\hrefCMSnoop {}{ {et~al.}, ``{Rivet user manual}'',} \textit{ Comput.
  Phys. Commun.} \textbf{ 184} (2013) 2803,
  \href{http://dx.doi.org/10.1016/j.cpc.2013.05.021}{\doi{10.1016/j.cpc.2013.05.021}},
\href{http://www.arXiv.org/abs/1003.0694}{\texttt{arXiv:1003.0694}}.

\bibitem{PDG2018}
\hrefCMSnoop {}{{Particle Data Group}, M.~Tanabashi {et~al.}, ``Review of
  particle physics'',} \textit{ Phys. Rev. D} \textbf{ 98} (2018) 030001,
  \href{http://dx.doi.org/10.1103/PhysRevD.98.030001}{\doi{10.1103/PhysRevD.98.030001}}.

\bibitem{Nussinov:1975ay}
\hrefCMSnoop {}{S.~Nussinov, ``{Possible effects of decays of charmed-particle
  resonances}'',} \textit{ Phys. Rev. Lett.} \textbf{ 35} (1975) 1672,
\href{http://dx.doi.org/10.1103/PhysRevLett.35.1672}{\doi{10.1103/PhysRevLett.35.1672}}.

\bibitem{Feldman:1977ir}
\hrefCMSnoop {}{G.~J. Feldman {et~al.}, ``{Observation of the decay $D^{*+} \to
  D^0 \pi^+$}'',} \textit{ Phys. Rev. Lett.} \textbf{ 38} (1977) 1313,
\href{http://dx.doi.org/10.1103/PhysRevLett.38.1313}{\doi{10.1103/PhysRevLett.38.1313}}.

\bibitem{Fruhwirth:1027031}
\hrefCMSnoop {}{R.~Fruhwirth, W.~Waltenberger, and P.~Vanlaer, ``{Adaptive
  vertex fitting}'',} \textit{ J. Phys.} \textbf{ G34} (2007) N343,
\href{http://dx.doi.org/10.1088/0954-3899/34/12/N01}{\doi{10.1088/0954-3899/34/12/N01}}.
“

\bibitem{Lisovyi:2015uqa}
\hrefCMSnoop {}{M.~Lisovyi, A.~Verbytskyi, and O.~Zenaiev, ``{Combined analysis
  of charm-quark fragmentation-fraction measurements}'',} \textit{ Eur. Phys.
  J. C} \textbf{ 76} (2016) 397,
  \href{http://dx.doi.org/10.1140/epjc/s10052-016-4246-y}{\doi{10.1140/epjc/s10052-016-4246-y}},
\href{http://www.arXiv.org/abs/1509.01061}{\texttt{arXiv:1509.01061}}.

\bibitem{CMS-PAS-LUM-17-001}
\hrefCMSnoop {}{{CMS Collaboration}, ``{CMS luminosity measurements for the
  2016 data taking Period}'',} CMS Physics Analysis Summary CMS-PAS-LUM-17-001,
  2017.
\newblock \url{http://cds.cern.ch/record/2257069}.

\bibitem{CMS-PAS-TRK-10-002}
\hrefCMSnoop {}{{CMS Collaboration}, ``{Measurement of tracking efficiency}'',}
  CMS Physics Analysis Summary CMS-PAS-TRK-10-002, 2010.
\newblock \url{http://cds.cern.ch/record/1279139}.

\bibitem{CrystalBallRef}
\href {http://www.slac.stanford.edu/cgi-wrap/getdoc/slac-r-236.pdf}{M.~J.
  Oreglia, ``A study of the reactions $\psi^\prime \to \gamma \gamma \psi$''}.
\newblock PhD thesis, Stanford University, 1980.
\newblock {SLAC} Report {SLAC-R-236}.

\bibitem{CMS-PAS-JME-16-004}
\hrefCMSnoop {}{{CMS Collaboration}, ``{Performance of missing energy
  reconstruction in 13 TeV pp collision data using the CMS detector}'',} CMS
  Physics Analysis Summary CMS-PAS-JME-16-004, 2016.
\newblock \url{https://cds.cern.ch/record/2205284}.

\bibitem{Aaron:2008ac}
\hrefCMSnoop {}{{H1} Collaboration, ``{Study of charm fragmentation into
  D$^{*\pm}$ mesons in deep-inelastic scattering at HERA}'',} \textit{ Eur.
  Phys. J. C} \textbf{ 59} (2009) 589,
  \href{http://dx.doi.org/10.1140/epjc/s10052-008-0792-2}{\doi{10.1140/epjc/s10052-008-0792-2}},
\href{http://www.arXiv.org/abs/0808.1003}{\texttt{arXiv:0808.1003}}.

\bibitem{Bowler:1981sb}
\hrefCMSnoop {}{M.~G. Bowler, ``{e$^+$e$^-$ Production of heavy quarks in the
  string model}'',} \textit{ Z. Phys. C} \textbf{ 11} (1981) 169,
\href{http://dx.doi.org/10.1007/BF01574001}{\doi{10.1007/BF01574001}}.

\bibitem{Andersson:1983ia}
\hrefCMSnoop {}{B.~Andersson, G.~Gustafson, G.~Ingelman, and T.~Sj{\"o}strand,
  ``{Parton fragmentation and string dynamics}'',} \textit{ Phys. Rept.}
  \textbf{ 97} (1983) 31,
\href{http://dx.doi.org/10.1016/0370-1573(83)90080-7}{\doi{10.1016/0370-1573(83)90080-7}}.

\bibitem{Alekhin:2018pai}
\hrefCMSnoop {}{S.~Alekhin, J.~Bl{\"u}mlein, and S.~Moch, ``{NLO PDFs from the
  ABMP16 fit}'',} \textit{ Eur. Phys. J. C} \textbf{ 78} (2018) 477,
  \href{http://dx.doi.org/10.1140/epjc/s10052-018-5947-1}{\doi{10.1140/epjc/s10052-018-5947-1}},
\href{http://www.arXiv.org/abs/1803.07537}{\texttt{arXiv:1803.07537}}.

\bibitem{Dulat:2015mca}
S.~Dulat\hrefCMSnoop {}{ {et~al.}, ``{New parton distribution functions from a
  global analysis of quantum chromodynamics}'',} \textit{ Phys. Rev. D}
  \textbf{ 93} (2016) 033006,
  \href{http://dx.doi.org/10.1103/PhysRevD.93.033006}{\doi{10.1103/PhysRevD.93.033006}},
\href{http://www.arXiv.org/abs/1506.07443}{\texttt{arXiv:1506.07443}}.

\bibitem{Harland-Lang:2014zoa}
\hrefCMSnoop {}{L.~A. Harland-Lang, A.~D. Martin, P.~Motylinski, and R.~S.
  Thorne, ``{Parton distributions in the LHC era: MMHT 2014 PDFs}'',} \textit{
  Eur. Phys. J. C} \textbf{ 75} (2015) 204,
  \href{http://dx.doi.org/10.1140/epjc/s10052-015-3397-6}{\doi{10.1140/epjc/s10052-015-3397-6}},
\href{http://www.arXiv.org/abs/1412.3989}{\texttt{arXiv:1412.3989}}.

\bibitem{Ball:2017nwa}
\hrefCMSnoop {}{{NNPDF} Collaboration, ``{Parton distributions from
  high-precision collider data}'',} \textit{ Eur. Phys. J. C} \textbf{ 77}
  (2017) 663,
  \href{http://dx.doi.org/10.1140/epjc/s10052-017-5199-5}{\doi{10.1140/epjc/s10052-017-5199-5}},
\href{http://www.arXiv.org/abs/1706.00428}{\texttt{arXiv:1706.00428}}.

\bibitem{Alekhin:2017kpj}
\hrefCMSnoop {}{S.~Alekhin, J.~Bl{\"u}mlein, S.~Moch, and R.~Pla{\v c}akyt{\.
  e}, ``{Parton distribution functions, $\alpha_s$, and heavy-quark masses for
  LHC Run II}'',} \textit{ Phys. Rev. D} \textbf{ 96} (2017) 014011,
  \href{http://dx.doi.org/10.1103/PhysRevD.96.014011}{\doi{10.1103/PhysRevD.96.014011}},
\href{http://www.arXiv.org/abs/1701.05838}{\texttt{arXiv:1701.05838}}.

\bibitem{Alekhin:2015cza}
\hrefCMSnoop {}{S.~Alekhin, J.~Bl{\"u}mlein, S.~Moch, and R.~Pla{\v c}akyt{\.
  e}, ``{Isospin asymmetry of quark distributions and implications for single
  top-quark production at the LHC}'',} \textit{ Phys. Rev. D} \textbf{ 94}
  (2016) 114038,
  \href{http://dx.doi.org/10.1103/PhysRevD.94.114038}{\doi{10.1103/PhysRevD.94.114038}},
\href{http://www.arXiv.org/abs/1508.07923}{\texttt{arXiv:1508.07923}}.

\bibitem{Aaron:2009aa}
\hrefCMSnoop {}{{H1 and ZEUS Collaborations}, ``{Combined measurement and QCD
  analysis of the inclusive $\Pe^\pm \Pp$ scattering cross sections at
  HERA}'',} \textit{ JHEP} \textbf{ 01} (2010) 109,
  \href{http://dx.doi.org/10.1007/JHEP01(2010)109}{\doi{10.1007/JHEP01(2010)109}},
\href{http://www.arXiv.org/abs/0911.0884}{\texttt{arXiv:0911.0884}}.

\bibitem{Abramowicz:2015mha}
\hrefCMSnoop {}{{H1 and ZEUS Collaborations}, ``{Combination of measurements of
  inclusive deep inelastic ${\Pe^{\pm }\Pp}$ scattering cross sections and QCD
  analysis of HERA data}'',} \textit{ Eur. Phys. J. C} \textbf{ 75} (2015) 580,
  \href{http://dx.doi.org/10.1140/epjc/s10052-015-3710-4}{\doi{10.1140/epjc/s10052-015-3710-4}},
\href{http://www.arXiv.org/abs/1506.06042}{\texttt{arXiv:1506.06042}}.

\bibitem{Khachatryan:2016pev}
\hrefCMSnoop {}{{CMS Collaboration}, ``{Measurement of the differential cross
  section and charge asymmetry for inclusive $\mathrm {p}\mathrm {p}\rightarrow
  \mathrm {W}^{\pm }+X$ production at ${\sqrt{s}} = 8$ TeV}'',} \textit{ Eur.
  Phys. J. C} \textbf{ 76} (2016) 469,
  \href{http://dx.doi.org/10.1140/epjc/s10052-016-4293-4}{\doi{10.1140/epjc/s10052-016-4293-4}},
\href{http://www.arXiv.org/abs/1603.01803}{\texttt{arXiv:1603.01803}}.

\bibitem{Carli:2010rw}
T.~Carli\hrefCMSnoop {}{ {et~al.}, ``{A posteriori inclusion of parton density
  functions in NLO QCD final-state calculations at hadron colliders: the
  APPLGRID project}'',} \textit{ Eur. Phys. J. C} \textbf{ 66} (2010) 503,
  \href{http://dx.doi.org/10.1140/epjc/s10052-010-1255-0}{\doi{10.1140/epjc/s10052-010-1255-0}},
\href{http://www.arXiv.org/abs/0911.2985}{\texttt{arXiv:0911.2985}}.

\bibitem{Alekhin:2014irh}
\hrefCMSnoop {}{S.~Alekhin {et~al.}, ``{HERAFitter}'',} \textit{ Eur. Phys. J.
  C} \textbf{ 75} (2015) 304,
  \href{http://dx.doi.org/10.1140/epjc/s10052-015-3480-z}{\doi{10.1140/epjc/s10052-015-3480-z}},
\href{http://www.arXiv.org/abs/1410.4412}{\texttt{arXiv:1410.4412}}.

\bibitem{herafitter}
\hrefCMSnoop {}{}2018.
\newblock {xFitter} web site, \url{http://www.xfitter.org/xFitter}.

\bibitem{Gribov:1972ri}
\hrefCMSnoop {}{V.~N. Gribov and L.~N. Lipatov, ``{Deep inelastic $\Pe$-$\Pp$
  scattering in perturbation theory}'',} \textit{ Sov. J. Nucl. Phys.} \textbf{
  15} (1972)
438.

\bibitem{Altarelli:1977zs}
\hrefCMSnoop {}{G.~Altarelli and G.~Parisi, ``{Asymptotic freedom in parton
  language}'',} \textit{ Nucl. Phys. B} \textbf{ 126} (1977) 298,
\href{http://dx.doi.org/10.1016/0550-3213(77)90384-4}{\doi{10.1016/0550-3213(77)90384-4}}.

\bibitem{Curci:1980uw}
\hrefCMSnoop {}{G.~Curci, W.~Furmanski, and R.~Petronzio, ``{Evolution of
  parton densities beyond leading order: the nonsinglet case}'',} \textit{
  Nucl. Phys. B} \textbf{ 175} (1980) 27,
\href{http://dx.doi.org/10.1016/0550-3213(80)90003-6}{\doi{10.1016/0550-3213(80)90003-6}}.

\bibitem{Furmanski:1980cm}
\hrefCMSnoop {}{W.~Furmanski and R.~Petronzio, ``{Singlet parton densities
  beyond leading order}'',} \textit{ Phys. Lett. B} \textbf{ 97} (1980) 437,
\href{http://dx.doi.org/10.1016/0370-2693(80)90636-X}{\doi{10.1016/0370-2693(80)90636-X}}.

\bibitem{Moch:2004pa}
\hrefCMSnoop {}{S.~Moch, J.~A.~M. Vermaseren, and A.~Vogt, ``{The three-loop
  splitting functions in QCD: the non-singlet case}'',} \textit{ Nucl. Phys. B}
  \textbf{ 688} (2004) 101,
  \href{http://dx.doi.org/10.1016/j.nuclphysb.2004.03.030}{\doi{10.1016/j.nuclphysb.2004.03.030}},
\href{http://www.arXiv.org/abs/hep-ph/0403192}{\texttt{arXiv:hep-ph/0403192}}.

\bibitem{Vogt:2004mw}
\hrefCMSnoop {}{A.~Vogt, S.~Moch, and J.~A.~M. Vermaseren, ``{The three-loop
  splitting functions in QCD: the singlet case}'',} \textit{ Nucl. Phys. B}
  \textbf{ 691} (2004) 129,
  \href{http://dx.doi.org/10.1016/j.nuclphysb.2004.04.024}{\doi{10.1016/j.nuclphysb.2004.04.024}},
\href{http://www.arXiv.org/abs/hep-ph/0404111}{\texttt{arXiv:hep-ph/0404111}}.

\bibitem{Botje:2010ay}
\hrefCMSnoop {}{M.~Botje, ``{QCDNUM: fast QCD evolution and convolution}'',}
  \textit{ Comput. Phys. Commun.} \textbf{ 182} (2011) 490,
  \href{http://dx.doi.org/10.1016/j.cpc.2010.10.020}{\doi{10.1016/j.cpc.2010.10.020}},
\href{http://www.arXiv.org/abs/1005.1481}{\texttt{arXiv:1005.1481}}.

\bibitem{Thorne:2006qt}
\hrefCMSnoop {}{R.~S. Thorne, ``{Variable-flavor number scheme for
  next-to-next-to-leading order}'',} \textit{ Phys. Rev. D} \textbf{ 73} (2006)
  054019,
  \href{http://dx.doi.org/10.1103/PhysRevD.73.054019}{\doi{10.1103/PhysRevD.73.054019}},
\href{http://www.arXiv.org/abs/hep-ph/0601245}{\texttt{arXiv:hep-ph/0601245}}.

\bibitem{Martin:2009ad}
\hrefCMSnoop {}{A.~D. Martin, W.~J. Stirling, R.~S. Thorne, and G.~Watt,
  ``{Parton distributions for the LHC}'',} \textit{ Eur. Phys. J. C} \textbf{
  63} (2009) 189,
  \href{http://dx.doi.org/10.1140/epjc/s10052-009-1072-5}{\doi{10.1140/epjc/s10052-009-1072-5}},
\href{http://www.arXiv.org/abs/0901.0002}{\texttt{arXiv:0901.0002}}.

\bibitem{Pumplin:2001ct}
J.~Pumplin\hrefCMSnoop {}{ {et~al.}, ``{Uncertainties of predictions from
  parton distribution functions. II. The Hessian method}'',} \textit{ Phys.
  Rev. D} \textbf{ 65} (2001) 014013,
  \href{http://dx.doi.org/10.1103/PhysRevD.65.014013}{\doi{10.1103/PhysRevD.65.014013}},
\href{http://www.arXiv.org/abs/hep-ph/0101032}{\texttt{arXiv:hep-ph/0101032}}.

\bibitem{Giele:1998gw}
\hrefCMSnoop {}{W.~T. Giele and S.~Keller, ``{Implications of hadron collider
  observables on parton distribution function uncertainties}'',} \textit{ Phys.
  Rev. D} \textbf{ 58} (1998) 094023,
  \href{http://dx.doi.org/10.1103/PhysRevD.58.094023}{\doi{10.1103/PhysRevD.58.094023}},
\href{http://www.arXiv.org/abs/hep-ph/9803393}{\texttt{arXiv:hep-ph/9803393}}.

\bibitem{Giele:2001mr}
\hrefCMSnoop {}{W.~T. Giele, S.~A. Keller, and D.~A. Kosower, ``{Parton
  distribution function uncertainties}'',} (2001).
\href{http://www.arXiv.org/abs/hep-ph/0104052}{\texttt{arXiv:hep-ph/0104052}}.

\end{thebibliography}\endgroup
